\documentclass[sigconf,authorversion,nonacm]{acmart}
\usepackage{balance}
\usepackage{tikz}
\usepackage[ruled,vlined,linesnumbered]{algorithm2e}
\setcopyright{none}

\title[Offline Risk-sensitive RL with Partial Observability]{Offline Risk-sensitive RL with Partial Observability to Enhance Performance in Human-Robot Teaming}

\author{Giorgio Angelotti}
\affiliation{
  \institution{ANITI, Fédération ENAC ISAE-SUPAERO ONERA,\\ Université de Toulouse, France}
  \city{}
  \country{}
  }
\email{giorgio.angelotti@isae-supaero.fr}

\author{Caroline P. C. Chanel}
\affiliation{
  \institution{ANITI, Fédération ENAC ISAE-SUPAERO ONERA,\\ Université de Toulouse, France}
  \city{}
  \country{}
  }
\email{caroline.chanel@isae-supaero.fr}

\author{Adam Henrique Moreira Pinto}
\affiliation{
  \institution{Fédération ENAC\\ ISAE-SUPAERO ONERA,\\ Université de Toulouse, France}
  \city{}
  \country{}
  }
\email{adam.moreira-pinto@isae-supaero.fr}

\author{Christophe Lounis}
\affiliation{
  \institution{Fédération ENAC\\ ISAE-SUPAERO ONERA,\\ Université de Toulouse, France}
  \city{}
  \country{}
  }
\email{christophe.lounis@isae-supaero.fr}

\author{Corentin Chauffaut}
\affiliation{
  \institution{Fédération ENAC\\ ISAE-SUPAERO ONERA,\\ Université de Toulouse, France}
  \city{}
  \country{}
  }
\email{corentin.chauffaut@isae-supaero.fr}

\author{Nicolas Drougard}
\affiliation{
  \institution{ANITI, Fédération ENAC ISAE-SUPAERO ONERA,\\ Université de Toulouse, France}
  \city{}
  \country{}
  }
\email{nicolas.drougard@isae-supaero.fr}

\begin{abstract}
The integration of physiological computing into mixed-initiative human-robot interaction systems offers valuable advantages in autonomous task allocation by incorporating real-time features as human state observations into the decision-making system. This approach may alleviate the cognitive load on human operators by intelligently allocating mission tasks between agents. Nevertheless, accommodating a diverse pool of human participants with varying physiological and behavioral measurements presents a substantial challenge. To address this, resorting to a probabilistic framework becomes necessary, given the inherent uncertainty and partial observability on the human's state.
Recent research suggests to learn a Partially Observable Markov Decision Process (POMDP) model from a data set of previously collected experiences that can be solved using Offline Reinforcement Learning (ORL) methods. In the present work, we not only highlight the potential of partially observable representations and physiological measurements to improve human operator state estimation and performance, but also enhance the overall mission effectiveness of a human-robot team. Importantly, as the fixed data set may not contain enough information to fully represent complex stochastic processes, we propose a method to incorporate model uncertainty, thus enabling risk-sensitive sequential decision-making.
Experiments were conducted with a group of twenty-six human participants within a simulated robot teleoperation environment, yielding empirical evidence of the method's efficacy.

The obtained adaptive task allocation policy led to statistically significant higher scores than the one that was used to collect the data set, allowing for generalization across diverse participants also taking into account risk-sensitive metrics.
\end{abstract}


\newcommand{\BibTeX}{\rm B\kern-.05em{\sc i\kern-.025em b}\kern-.08em\TeX}

\begin{document}

\pagestyle{fancy}
\fancyhead{}

\maketitle 

\section{Introduction}
Over the past few years, Artificial Intelligence (AI) technologies have become widespread, playing a crucial role in automation across various industries and improving efficiency and productivity \cite{hall2017growing}. Hand in hand with the diffusion of AI, the need for regulations has been added to the agendas of law- and policy-makers. For instance, the recent European Union (EU) AI Act \cite{european2020white} promotes responsible AI deployment, emphasizing the need for robustness, interpretability, and human supervision, particularly relevant in Reinforcement Learning (RL).
The integration of humans in the workflow of Machine Learning (ML) models has been conceived in multiple ways \cite{mosqueira2023human}. This field, called Human-In-The-Loop (HITL) ML \cite{mosqueira2023human}, has been deemed necessary across several applications that require high-quality output and partial or total human supervision.

In this work, we focus on a specific niche of HITL-ML: HITL-RL and in particular HITL Offline RL (HITL-ORL). Offline RL can be seen as a completely separate branch of the RL community, aiming to obtaining an optimal control policy for an agent in an offline setting using a fixed batch of demonstrations. It is particularly well-suited to settings where the AI agent interacts with humans. In these scenarios, running parallelized and massive amount of simulations to train the agents is difficult or infeasible, since 
data collection involving human beings can be expensive, time-consuming, and possibly dangerous. 
Note that in the HITL-ORL context we consider, learning is not supervised by the human in the loop \cite{akkaladevi2018toward}. On the contrary, the human presence (\textit{i.e.} the human included in the system) \cite{singh2022pomdp,nikolaidis2015efficient} provides additional uncertain and partial observable features - as some relevant information about the human is not directly observable \cite{roy2020can} - and poses significant difficulties for learning a policy with ORL methods. We specifically consider the context where ORL is used to obtain a control policy to drive the allocation of tasks between agents \cite{hearst1999mixed,javdani2015shared,schilling2019shared}. 
The control system we aim to obtain should compute adaptive control policies, taking into account not only mission-related markers but also the state and behavior of the human operator.

Partially Observable Markov Decision Processes (POMDPs) have been used to make decisions in Human-Robot Interaction (HRI) \cite{goodrich2008human} to deal with environment uncertainty and partial observability \cite{nikolaidis2015efficient,nikolaidis2017human,gateau2016considering}. POMDPs allow for principled decision-making when human behavior is involved, including scenarios that involve multimodal observations (speech, eye gaze, and pointing gestures) \cite{nikolaidis2015efficient,nikolaidis2017game,nikolaidis2017human,jain2019probabilistic,singh2022pomdp,rosen2020mixed}. POMDP-based approaches are applied in various human-machine interaction cases, including robotics \cite{nikolaidis2015efficient,nikolaidis2017human,lauri2022partially}, medical diagnosis \cite{zhang2022diagnostic}, and assisting students \cite{taha2011pomdp}. POMDP models in HRI are typically specified by experts \cite{gateau2016considering}, either fully or partially. Some research explores the possibility of robots learning parts of the POMDP model from user interactions, even in cases with unknown reward functions or observation models \cite{nikolaidis2015efficient,singh2022pomdp}. However, as far as we know, no work has addressed in a unified manner the challenge of learning POMDP models from a fixed and limited data set, and the challenge of computing robust policies to the resulting (possibly inaccurate) models.

As previously stated, ORL aims to address the particular case when AI agents interact with the environment in contexts where trial-and-error learning could result in disastrous consequences \cite{levine2020offline,prudencio2023survey}. In ORL, the learning phase leverages a pre-collected data set. As discussed, the presence of humans, as part of the system, impacts the initial data set's informativeness in several ways: increased environmental stochasticity due to unpredictable behavior, limited data set size, and augmented biases. 
Addressing these challenges necessitates advanced ORL algorithms, such as \cite{lobo2021softrobust,yu2020mopo,kidambi2020morel,kumar2020conservative}. Interestingly, these model-based or model-free algorithms mainly focus on computing risk-sensitive or robust policies that could handle model uncertainties. 

Unfortunately, the state-of-the-art ORL algorithms \cite{kumar2019stabilizing,petrik2019beyond,yu2020mopo,kidambi2020morel,behzadian2021optimizing,lobo2021softrobust} were not designed to work with partially observable environments. Recently, \citet{janner2021offline} and \citet{chen2021decision} propose Transformer-based Deep Neural Network architectures that also deal with partial observable scenarios, but most Transformer-based methods do not perform at their best when confronted with limited data sets \cite{wang2022bootstrapped}. 

On top of this, which algorithm should a practitioner use among the many available ones? Interestingly, \citet{angelotti2022an} proposed Exploitation vs Caution (EvC), a method for offline risk-sensitive policy selection in low-dimensional Markov Decision Processes (MDPs) that resorts to a Bayesian estimate of model uncertainty starting from a fixed and restricted set of previously collected experiences. EvC selects the most performing policy in a set of candidate ones by computing offline policy evaluation in a Monte Carlo fashion. By exploiting a Bayesian representation of model uncertainty, several models are sampled from the Bayesian posterior and every policy in the candidate set is evaluated on those models until the estimate of a quantile of the distribution of performance (according to model uncertainty) falls within a desired confidence interval. Then, the best policy according to a risk-sensitive metric is selected. Although this method does not scale for high-dimensional decision processes, it seems promising to adapt it to select risk-sensitive policies for data-driven POMDPs, thereby considering model uncertainty and including partial observability, in a data-frugal regime. In this sense, we address in a unified way the challenge of learning a POMDP model and of computing robust policies for the resulting (possibly inaccurate) model.
In this exciting context, the contributions of the present work are the following:
\begin{itemize}
\item We propose a methodology to approximate a POMDP model including a Bayesian representation of model uncertainty, and we implement it through a problem-specific pipeline representing a Human-Robot Interaction study case. 
In detail, the starting limited data set is the result of eighteen human operators who interacted with the robotic system in laboratory settings using a random interaction policy.  
Using this data set, the Human-Robot system is modeled as a POMDP, achieving an interpretable state space representation and a Bayesian representation of model uncertainty. 

\item We extend the EvC policy selection method to consider model uncertainty in partially observable domains. To do this, we first compute different policies for the data-driven POMDP model. Then, by resorting to the Bayesian formalism and Monte Carlo sampling, we extend the EvC method to handle partial observable domains and to select the safest policy according to a risk-sensitive measure and POMDP model uncertainty.

\item We compare the robustness and the performance of the obtained robust POMDP policy with that used to collect the original data set, among others. 
This comparison is done by performing a novel set of experiments involving twenty-six human operators in laboratory facilities. 
The obtained robust POMDP policy led to statistically significant higher scores than the one that was used to collect the data set, allowing for generalization across diverse participants.
\end{itemize}

The paper is organized as follows: in the next section, we briefly review the technical background and related works. Then, in Section \ref{sec:frg_desc}, we present the environment used for data collection and evaluation. In Section \ref{sec:method}, the applied methodology is described, and in Section \ref{sec:results}, we present our validation experiments and results. The paper concludes with a discussion and future work perspectives.

\section{Background and Related works}
\paragraph{(PO)MDP Framework} Markov Decision Processes (MDPs) are a general mathematical framework that can describe the discrete-time progression of a Markovian stochastic process, while the actions of an agent can impact its evolution \cite{mausam2012planning}. Partially Observable Markov Decision Processes (POMDPs) are an extension of the MDP framework in which the agent has access only to partial observations of the state of the environment \cite{kurniawati2008sarsop}. In both cases, the agent receives a reward signal $r_t \in \mathbb{R}$ from the environment at each time step. Note that a Hidden Markov Model (HMM) can be seen as a particular POMDP case in which the state transition function does not depend on the actions and the reward function is not defined. Usually, solving a (PO)MDP means computing a policy, \textit{i.e.} a function or a set of rules, that allows 
the agent to take actions to maximize the discounted expected cumulative reward obtained along a possible trajectory, $\mathbb{E}[\sum_{t=0}^{\infty}\gamma^t r_t]$, where $0 < \gamma < 1$ is the discount factor. The agent can compute such a policy by \textit{solving} a (PO)MDP. However, solving a POMDP for an infinite time horizon is undecidable \cite{madani1999undecidability}, and usually only approximate solutions can be obtained in reasonable time \cite{kurniawati2008sarsop}.

\paragraph{Offline POMDP model learning and solving}
When partial observability enters the scene, learning a model from demonstrations becomes an exceedingly complicated task. Nevertheless, offline POMDP learning is of great importance as a representative model of the system dynamics might be impossible to be written by hand, especially when there are humans in the loop. An important issue in offline POMDP learning is intrinsic and rooted at the foundation of any possible learning procedure: which representation is the most appropriate? Remember that in a POMDP, the agent has only access to \textit{observations}, while the system evolves over time by transitioning across changing, partially observable states. To implement any learning algorithm, one must first deal with the choice of how to represent the (hidden) state space. 
A second main issue is model uncertainty. For instance, in the case of a discrete MDP, it is relatively straightforward to include uncertainty about the model estimate within a Dirichlet distribution, initialized with the frequency of transitions in the data set. However, such convenient modeling is not available when dealing with POMDPs. Moreover, one should account for not only the uncertainty about the transition function but also the observation function. Offline POMDP learning is a problem far more difficult than offline MDP learning.

Offline POMDP learning has been investigated by the research community, but to the best of our knowledge, no one has managed to develop a general approach. The majority of existing methods that learn a POMDP offline to address planning for robotics define the representation with expert guidance. In the works from \cite{atrash2010bayesian,taha2011pomdp,gopalan2015modeling}, the POMDP representation is provided by an oracle, and only transition and observation functions are learned. The work in \cite{broz2011designing} assumes that the states are described by discrete variables and the observations can be just a subset of the latter. In the context where the histories composing the data set have been generated by an unknown POMDP, the work in \cite{franccois2019overfitting} shows that choosing a discrete state space with a lower dimensionality than the original one can help reduce overfitting and hence yield relevant policies. Interestingly, the same work also advocates that reducing the discount factor during planning can further benefit the performance of the policy at the time of deployment.
Since learning a POMDP model from data is already challenging, model uncertainty is a difficulty that very few approaches have attempted to include and address. The works in \cite{atrash2010bayesian,doshi2012reinforcement} acknowledge that a learned trivial model might not be representative enough. For this purpose, they established a Bayesian framework from which models are extracted by a Bayesian prior. However, model uncertainty is then reduced in an online setting by allowing the agent to interact with an oracle.

In this context, and inspired by the Bayesian framework proposed in \cite{atrash2010bayesian,doshi2012reinforcement}, we believe that extending
a method for offline risk-sensitive policy selection resorting to a Bayesian estimate of model uncertainty, 
namely Exploitation vs Caution \cite{angelotti2022an},
to POMDPs, could be useful for selecting the most risk-sensitive policy among those obtained using different solvers and/or varying hyperparameters - e.g. $\gamma$ as suggested by \citet{franccois2019overfitting}. Indeed, in a (PO)MDP, $1-\gamma$ can be interpreted as the time-step probability to exit from the modeled stochastic process \cite{lattimore2014general}. \citet{jiang2015dependence} shows that in the context of data-driven model learning there is an optimal discount factor
$\gamma^* < \gamma < 1$
that should be used to compute a policy for eventual real-world deployment. In light of this, our approach aims to use EvC and select a policy computed with a discount factor estimated to be closer to $\gamma^*$.

\section{The Firefighter Robot Game}
\label{sec:frg_desc}
The environment under study consists of a serious game, which can be played at \href{http://robot-isae.isae.fr}{\url{robot-isae.isae.fr}},  designed to generate deleterious cognitive states in human operators during the missions they carry out with an artificial agent, mainly due to multitasking, uncertainty, and time pressure. Such situations are known to generate stress and cognitive workload, thus impacting human agent performance \cite{drougard2017mixed,drougard2018physiological,dehais2020neuroergonomics}. 
Deleterious mental states affecting performance can be estimated using physiological computing \cite{roy2020can,fairclough2009fundamentals}. For instance, metrics such as the Heart Rate (HR) and the Heart Rate Variability (HRV) are known to be impacted by workload \cite{heard2018survey}. With this in mind, we believe that exploring human behavioral and physiological features in the construction and evaluation of human-system interaction policies is particularly promising. In this context, our goal is to use ORL to develop a robust policy for an interaction controller within a quasi-mixed initiative interaction system - the system adapts its behavior based on several mission parameters, including those related to the human operator.

\paragraph{Brief environment description} The Firefighter Robot Game (FRG) \cite{drougard2017mixed,charles2018human, chanel2020mixed} 
is a scenario where a human teleoperates a robot to extinguish fires in a confined area. The forest has nine trees that can catch fire, and the human-robot team must put out as many fires as possible within 10 minutes. The robot has limited battery power, a water tank for extinguishing fires, and a thermometer to monitor its temperature. The robot must recharge at an energy supply zone and refill its water tank at a water pool. However, the pool's walls are susceptible to leaks and necessitate manual intervention for mending. Additionally, filling the pool requires precise control of the movement of a moving filling nozzle.

An automated controller manages the human-robot interaction, switching between automatic and manual modes and activating or deactivating alarm notifications. In automatic mode, the robot prioritizes battery recharging and embedded water tank refilling. It navigates to the energy supplier or water pool when resources are low. However, the human operator is required to manage, repair, and refill the water pool. When battery and tank levels are sufficient, the robot finds the shortest path to the nearest burning tree and extinguishes the fire. In manual mode, the human operator remotely controls all robot actions, including navigation, recharging, temperature monitoring, and water dispensing.

\paragraph{Data collection}
Data were collected in a former study involving 18 participants \cite{chanel2020towards}. 
These participants underwent four missions, preceded by electrocardiogram (ECG) setup and eye-tracking calibration. 
Each mission was preceded by a rest period and followed by the filling of surveys and a break. 
During missions, the robot mode (automatic or manual) changed randomly every 10 seconds, as did the activation of the alarm system.
Metrics such as robot parameters, mission states, operator actions, ECG and eye-tracking data were recorded. 
Artifacts were removed from live ECG data to compute the HR and HRV measurements. Rest data helped to compute normalized HR and HRV metrics for each mission and participant. Eye-tracking data consisted of the number and duration of fixations in five areas of interest (see \cite{chanel2020towards} for more details). Human actions on the interface, such as keystrokes and clicks, were also recorded.
These characteristics were computed using 10-second intervals,
and associated with the current robot's automation mode, and status of the alarm system.
Henceforth, this data set will be denoted as $\mathcal{D}$.
The success of a mission is quantified by the number of fires extinguished, overlooking other human-centric metrics.

\section{Risk-sensitive ORL with partial observability}
\label{sec:method}
\begin{figure}[b]
    \centering
\begin{tikzpicture}
\node[rotate=90] at (-4.7,5.8) {\textbf{Batch of}};
\node[rotate=90] at (-4.3,5.8) {\textbf{experiences}};
\node[rotate=90] at (-3.9,5.8) {\textbf{($\mathcal{D}$)}};
\node at (-2.1,5.9) {\includegraphics[width=0.3\linewidth]{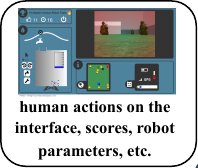}};
\node at (2.4,5.7) {\includegraphics[width=0.2\linewidth]{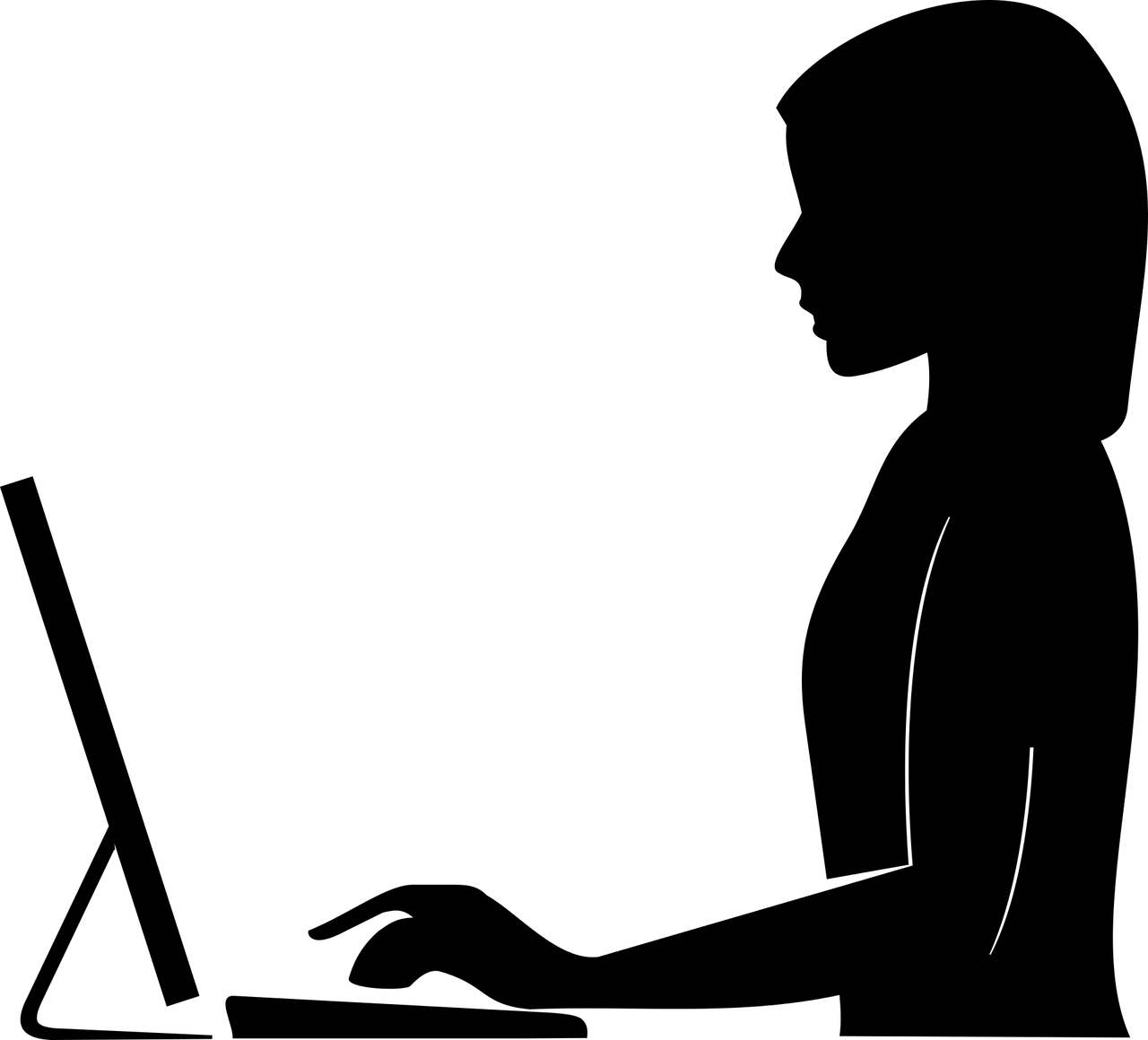}};
\node at (-0.3,6.5) {\includegraphics[width=0.09\linewidth]{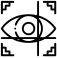}};
\node at (-0.3,5.65) {\includegraphics[width=0.11\linewidth]{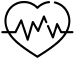}};
\draw[-latex, opacity=0.8,very thick] (0.2,6.5) -- (2.6,6.2);
\draw[-latex, opacity=0.8,very thick] (0.2,5.65) to [bend left] (2.5,5.7);
\draw[-latex, opacity=0.8,very thick] (-0.75,5.1) -- (1.5,5.3);
\draw[opacity=0.8,very thick, dashed] (-4.7,4.75) -- (3.3,4.75);
\node[rotate=90] at (-4.7,3) {\textbf{Risk-sensitive ORL}};
\node[rotate=90] at (-4.3,3) {\textbf{with partial}};
\node[rotate=90] at (-3.9,3) {\textbf{observability}};
\node at (0,3) {\includegraphics[width=0.8\linewidth]{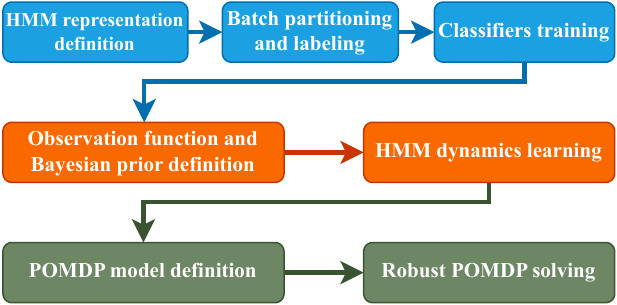}};
\draw[opacity=0.8,very thick, dashed] (-4.7,1.25) -- (3.3,1.25);
\node[rotate=90] at (-4.7,0) {\textbf{Validation}};
\node[rotate=90] at (-4.3,0) {\textbf{experiments}};
\node[rotate=90] at (-3.9,0) {\textbf{($\mathcal{D}_{val}$)}};
\node at (0,0) {\includegraphics[width=0.8\linewidth]{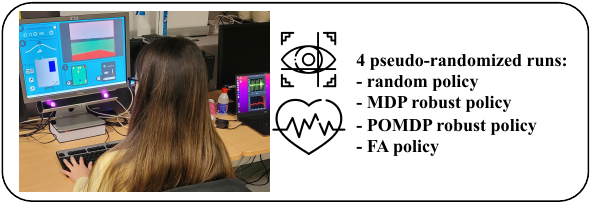}};
\draw[-latex, opacity=0.8,ultra thick] (-3.6,6.9) -- (-3.6,-1.1);
\end{tikzpicture}
\caption{Illustration of the proposed methodology.}
\Description{The proposed methodology is represented chronologically in three blocks from top to bottom. The first block represents the batch of past experiments, containing data on the operator's heart rate, gaze and interaction with the interface. The second block corresponds to risk-sensitive Offline Reinforcement Learning with partial observability and describes the learning, modeling and optimization steps leading to a robust interaction strategy, in the form of sub-blocks whose order is represented by arrows. Finally, the third block concerns validation experiments, and contains a photo of a participant carrying out a mission, as well as a list of the strategies tested.}
\label{fig:pipeline}
\end{figure}
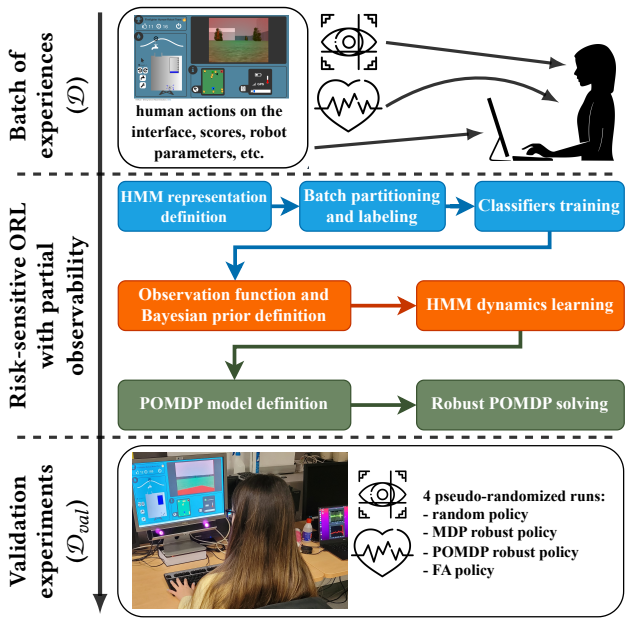

The Figure \ref{fig:pipeline} illustrates the proposed pipeline. The data set $\mathcal{D}$ was used as input for ORL. We adopted a model-based approach, such as using a POMDP representation, for its superior interpretability. The POMDP's hidden states marked team performance, while the visible ones indicated robot autonomy mode and alarm status. We considered four actions: \textit{put-manual-alarms-on}, \textit{put-manual-alarms-off}, \textit{put-automatic-alarms-on}, \textit{put-automatic-alarms-off}. For the partial observable case, we considered an observation set identical to the state set as suggested by \cite{broz2011designing,franccois2019overfitting}. Note that the observability concerning robot mode and alarm status is complete, although the observability concerning the current human-robot team performance is - by definition - partial. To learn the dynamics of such a model and the respective observation function, the methodology applied is hereafter described.

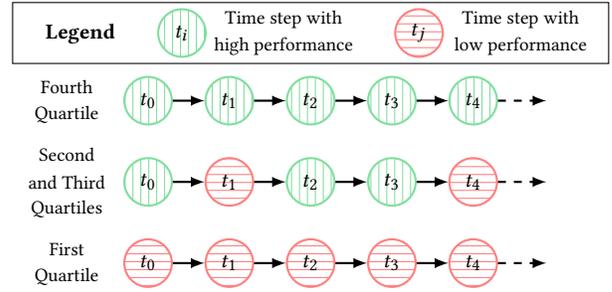
\begin{figure}[!tb]
\centering
\begin{tikzpicture}[transform shape,scale=0.9]
\usetikzlibrary {patterns,patterns.meta} 
\definecolor{darkpastelgreen}{rgb}{0.01, 0.75, 0.24}
\tikzstyle{rvertex}=[circle,pattern={horizontal lines},pattern color=red!50,minimum size=20pt,inner sep=0pt, draw=red!50,thick]
\tikzstyle{gvertex}=[circle,pattern={vertical lines},pattern color=darkpastelgreen!50,minimum size=20pt,inner sep=0pt, draw=darkpastelgreen!50,thick]

\node at (-1.2,0.2) {\small First};
\node at (-1.2,-0.2) {\small Quartile};
\foreach \name/\x in {t_0/0,t_1/1.2,t_2/2.4, t_3/3.6,t_4/4.8}
\node[rvertex] (G-\name) at (\x,0) {$\name$};
\node (G-end) at (6,0) {};
\foreach \from/\to in {t_0/t_1,t_1/t_2,t_2/t_3,t_3/t_4}
\draw[->,>=latex,thick] (G-\from) -- (G-\to);
\foreach \from/\to in {t_4/end}
\draw[->,>=latex,dashed,thick] (G-\from) -- (G-\to);

\def\stq{1.2}
\node at (-1.2,\stq+0.4) {\small Second};
\node at (-1.2,\stq) {\small and Third};
\node at (-1.2,\stq-0.4) {\small Quartiles};
\foreach \name/\x in {t_1/1.2,t_4/4.8}
\node[rvertex] (G-\name) at (\x,\stq) {$\name$};
\foreach \name/\x in {t_0/0,t_2/2.4, t_3/3.6}
\node[gvertex] (G-\name) at (\x,\stq) {$\name$};
\node (G-end) at (6,\stq) {};
\foreach \from/\to in {t_0/t_1,t_1/t_2,t_2/t_3,t_3/t_4}
\draw[->,>=latex,thick] (G-\from) -- (G-\to);
\foreach \from/\to in {t_4/end}
\draw[->,>=latex,dashed,thick] (G-\from) -- (G-\to);

\def\fq{2.4}
\node at (-1.2,\fq+0.2) {\small Fourth};
\node at (-1.2,\fq-0.2) {\small Quartile};
\foreach \name/\x in {t_0/0,t_1/1.2,t_2/2.4, t_3/3.6,t_4/4.8}
\node[gvertex] (G-\name) at (\x,\fq) {$\name$};
\node (G-end) at (6,\fq) {};
\foreach \from/\to in {t_0/t_1,t_1/t_2,t_2/t_3,t_3/t_4}
\draw[->,>=latex,thick] (G-\from) -- (G-\to);
\foreach \from/\to in {t_4/end}
\draw[->,>=latex,dashed,thick] (G-\from) -- (G-\to);

\def\lg{3.4}
\node at (-1,\lg) {\textbf{Legend}};
\node[gvertex] at (0.5,\lg) {$t_i$};
\node at (2,\lg+0.2) {\small Time step with};
\node at (2,\lg-0.2) {\small high performance};
\node[rvertex] at (4,\lg) {$t_j$};
\node at (5.5,\lg+0.2) {\small Time step with};
\node at (5.5,\lg-0.2) {\small low performance};
\draw (-2,\lg-0.45) rectangle (6.8,\lg+0.45);
\end{tikzpicture}
\caption{Example of how the batch is split.}
\label{fig:split}
\Description{The time steps of past experiences contained in the dataset are represented as circles distributed from left to right, with arrows representing the transition from one step to the next. The circles are green when performance is good, and red otherwise. These past experiences are ranked according to their final score, from best (top) to worst (bottom).  There are therefore almost exclusively good performances (green circles) in the fourth quartile (top), and bad performances (red circles) in the first quartile (bottom), but this is not the case for past experiences with intermediate scores (middle).}
\end{figure}

\paragraph{Data split.} The mission score in $\mathcal{D}$, which represents the total number of extinguished fires throughout an entire mission, guided data splitting and classifier training to map high-dimensional observations (raw recorded data) into a low-dimensional variable, whose two possible entries are \textit{performant} and \textit{non-performant}. Specifically, the batch $\mathcal{D}$ consists of $72$ mission recordings, each containing several feature vectors labeled by the mission's global score. However, to infer the system state variable concerning performance (including the mental engagement of a human operator), we need to evaluate a local measure of efficiency. To do this, we assume that during missions with a low score, the team exhibits constant non-performant behavior and, conversely, during missions with a very high score the team exhibits constant performant behavior. Hence, the time steps of the missions in the First Quartile (with respect to the final score) are labeled as \textit{non-performant} (red in Fig. \ref{fig:split}) and the ones in the top $25\%$ (Fourth Quartile) as \textit{performant} (green). The time steps in the missions in the second and third quartiles form the batch that will be used to learn the transition dynamics, as we assume that there might be frequent changes in performance status during missions with an average score.

\paragraph{Classifier training}
Using the data set partitioned and labeled as mentioned above, we train four different classifiers. These classifiers provide an observation of the mission performance for the four possible configurations of the two remaining visible state variables, namely the robot autonomy mode (autonomous/manual) and the alarm system status (on/off). Each trained classifier translates the features, related to time steps with one of the four visible state configurations, into a (noisy) observation of the hidden variable, namely mission performance.

We opted to train four Extra Tree Classifiers as indicated by \citet{geurts2006extremely}. We preferred Extra Tree Classifiers to Random Forests because the first produce results with reduced variance, even though with a slight increase in bias. This decision fits well with our objective, given that we aim for classifiers that can effectively generalize across various participants who may present a varied distribution in their measurements.
The classifiers are trained using a 10-fold Group (participant-wise) Shuffle Split cross-validation approach to select the best hyperparameters with respect to the balanced accuracy metric, initially using a Random Search and then with a Grid Search method. The validation set contains $20$\% of the data from the entire training set, and both sets include only missions carried out by different groups of participants. In doing so, we obtain the hyperparameters of the Extra Tree Classifiers that, on average, minimize the generalization error on the validation set.

\paragraph{Observation Function and Bayesian prior} The EvC paradigm requires a Bayesian representation of model uncertainty.
We note that the diagonal elements of the confusion matrix associated with each classifier enumerate the number of samples correctly classified by the respective algorithm. Conversely, the off-diagonal elements represent the number of samples classified as pertaining to a particular class (as determined by the column index), while their actual label belongs to a different class (as determined by the row index). Hence, we propose using the confusion matrices of the classifiers to compute Bayesian Dirichlet posteriors for an HMM observation function. Moreover, normalizing a confusion matrix by row yields another matrix whose elements are $O_{ij} \approx \text{Pr}\left(O_t = o_j | S_t = s_i\right)$. 
These last row-normalized confusion matrices could also serve as the observation function for defining an HMM process. By defining an observation function - either by sampling one from the posteriors or simply by row-normalizing the prior - we can learn the dynamics of an HMM via Expectation Maximization (EM) \cite{bilmes1998gentle}. The latter will be used to learn the dynamics of a POMDP model. Indeed, the uniform randomness of the data collection policy allows the HMM to be transformed into a POMDP.

\paragraph{POMDP model definition} We define the POMDP model, called the \textit{trivial POMDP}, that we will use both to obtain the robust policy and to deploy the policy at execution time during experiments with new human participants. Indeed, we will need this model to update the belief in real-time. The trivial POMDP is obtained by using the row-normalized confusion matrices as observation function and the subsequent transformed HMM dynamics inferred via EM. As reward function, we use the average number of fires extinguished by the human-robot team in the original data-set for each state.

\paragraph{Robust POMDP solving.} Inspired by the works in \cite{jiang2015dependence,franccois2019overfitting}, the trivial POMDP was solved using SARSOP \cite{kurniawati2008sarsop} with different discount factors, producing diverse policies (Lines 2-4, Algorithm \ref{algo:pseudo-evc}). We then adapted the EvC algorithm for risk-aware policy selection in the POMDP context. Non-normalized confusion matrices updated Dirichlet priors expressing observation model uncertainty. Diverse observation functions were sampled from these posteriors (Line 6), yielding different HMM dynamics (Line 7) and POMDPs (Line 8). Candidate policies were evaluated with these POMDPs (Lines 9-14), with beliefs updated based on the solved model. The Value-at-Risk \cite{rockafellar2002conditional} at risk level $q$, \textit{i.e.} the value of the q-order quantile, set to $0.5$, guided robust policy selection (Line 15). As a limitation of the approach, we do not continue sampling models until the estimates of the quantile for each policy falls within a confidence interval, as in \cite{angelotti2022an}, due to computational time constraints. Nevertheless, although lacking theoretical guarantees, this method seems to produce decent results in the tested scenario.

\begin{algorithm}[h]
\caption{Robust POMDP solving and Risk-sensitive Policy Selection}
\label{algo:pseudo-evc}
\DontPrintSemicolon
\KwIn{Trivial POMDP, $\Gamma$ set of discount factors, $\mathcal{D}$ batch of trajectories, $\mathrm{N}_M$ number of models to sample/learn, $\mathrm{N}_E$ number of histories per model, $q$ quantile's order}
\textbf{Initialization:} $M \gets \varnothing$ (empty list), $\Pi \gets \varnothing$ (empty list)\\
\ForAll{$\gamma$ in $\Gamma$}
    {
        $\pi \gets \text{Solve(Trivial POMDP, } \gamma)$\\
        Append $\pi$ to $\Pi$
    }
    
\For{$i$ from $1$ to $\mathrm{N}_M$}
    {
        Sample observation functions from the Dirichlet posterior distributions\\
        Learn transition functions using EM(obs. functions, $\mathcal{D}$)\\
        Append new POMDP model to $M$
    }
\ForAll{$\pi$ in $\Pi$}
    {
    $G_{\pi} \gets \varnothing$ (empty list)\\
        \ForAll{$pomdp$ in $M$}
            {
            	  \For{$i$ from $1$ to $\mathrm{N}_E$}
            	  	{
                	$R \gets \text{Total reward of a generated trajectory}$\\
                	Append $R$ to $G_{\pi}$
                	}
            }
    }
$\pi^* \gets \arg\max_{\pi \in \Pi} \text{VaR}_{q}[G_{\pi}]$\\
\Return $\pi^*$
\end{algorithm}

\section{Validation experiments}
\label{sec:results}
By performing validation experiments in the same environment (i.e. the Firefigther Robot Game) with different participants than those gathered during the initial data collection stage, the robustness and transferability of the policies can be fairly assessed. Preliminary power analysis was performed, requiring at least $30$ participants to perform a significant Friedman test with four groups at $0.95$ power (four policies will be compared).
Participants were incrementally recruited through electronic mail announcements and oral advertising, and recruitment stopped when the experiment's estimated power reached $0.95$. In the end, $26$ participants participated in the experiments (mean age $28.6$, sd. $5.7$; $7$ females). The experiments were approved by our local ethics committee. 
From now on, the data set including the validation experiments will be referred as $\mathcal{D}_{val}$. Fig. \ref{fig:pipeline} shows a participant taking part in the experiment. 

\subsection{Protocol}
The introductory phase of the experimental procedure included collecting informed consent, responsibility signatures, providing explanations, calibrating physiological measurement devices, preliminary training, and administering the Karolinska Sleepiness Scale (KSS) \cite{aakerstedt1990subjective} questionnaire. KSS is generally used to measure the participants' subjective fatigue; a higher post-experiment KSS score can indicate significant drowsiness. 
The core procedure of the experiments involved four missions in the FRG environment, each applying a distinct interaction control policy. These policies were pseudo-randomized among participants. Each run commenced with a baseline recording of cardiac activity (see Fig. \ref{fig:rest_hrv}), then proceeded to mission execution. Average HR and HRV during a 1-minute resting period were used as baselines to normalize subsequent HR and HRV values. Artifacts in HR and HRV measurements were removed in real-time. Participants then completed the NASA-TLX questionnaire \cite{hart1988development}, providing feedback on the perceived effort generated by the mission. Subsequently, they filled out a questionnaire inspired by the work in \cite{guy2019eval}, to provide feedback on the fluency of the current adaptive policy before taking a 2-minute break. 

\begin{figure*}[!h]
\centering
\begin{tikzpicture}
\node at (0,0) {\includegraphics[width=0.45\textwidth]{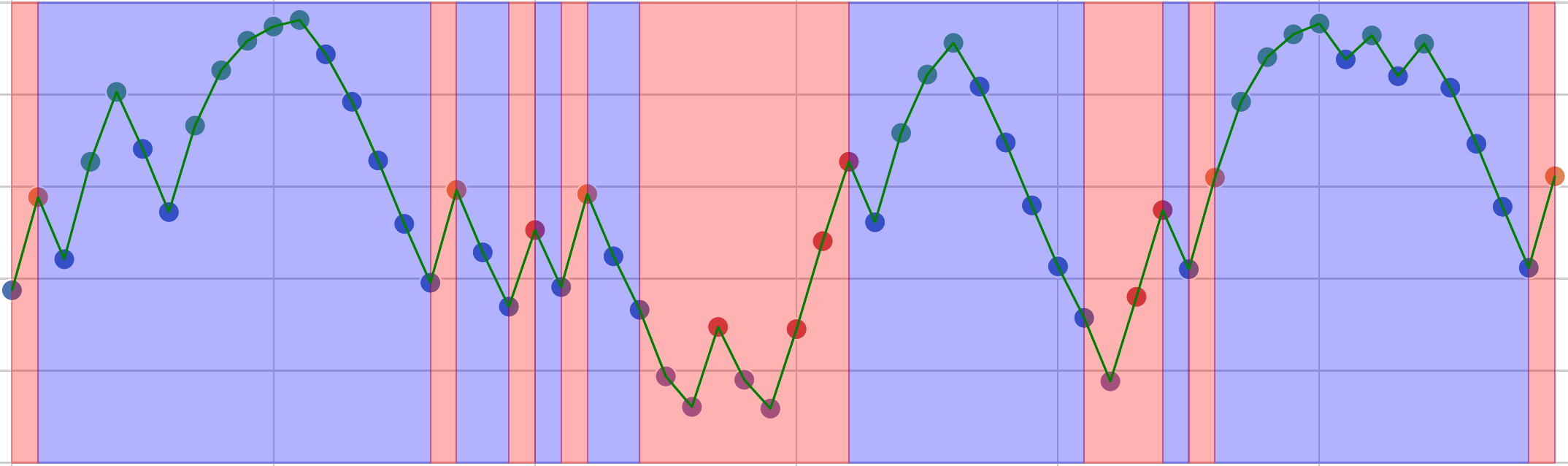}};
\node at (6,0) {\includegraphics[width=0.09\textwidth]{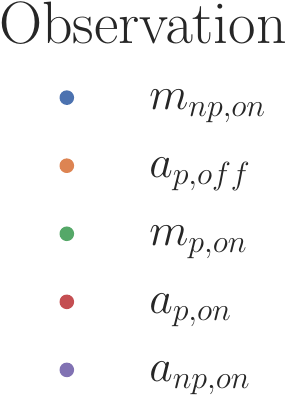}};
\node at (-6.2,0) {\includegraphics[width=0.085\textwidth]{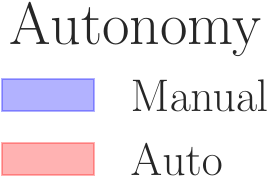}};
\draw[-latex, opacity=0.5] (-4,-1.17) -- (4.2,-1.17);
\node[rotate=90] at (-4.6,0) {\footnotesize Belief of Performance};
\foreach \name/\x in {0.0/0, 0.2/1, 0.4/2, 0.6/3, 0.8/4, 1.0/5}
\node () at (-4.2,-1.1+0.44*\x) {\footnotesize $\name$};
\foreach \name/\x in {100/1, 200/2, 300/3, 400/4, 500/5}
\node () at (-4.15+1.335*\x,-1.05) {\scriptsize $\name$};
\foreach \name/\x in {100/1, 200/2, 300/3, 400/4, 500/5}
\draw[opacity=0.5] (-3.95+1.335*\x,-1.17) -- (-3.95+1.335*\x,1.17);
\node () at (4.1, -1) {\footnotesize Time (sec)};
\end{tikzpicture}
\caption{Computation of POMDP's marginalized belief of performance $\beta_t$ along a mission run using the POMDP policy. The control system switches the autonomy mode of the robot from manual to automatic (auto), and vice-versa. 
Notice how the belief is different from the immediate observation, as multiple subsequent \textit{non-performant} observations can be necessary for the system to deem the state necessary to put the robot in automatic mode, e.g. from $t \approx 100 s$ to $t \approx 150 s$.}
\label{fig:time_belief}
\Description{The x-axis represents time (between 0 and 600), and the y-axis represents belief in performance (between 0 and 1). The curve therefore represents the evolution of belief over time for a mission in the validation dataset. The periods during which the robot is in its autonomous mode are represented by a red background, and by a blue background in its manual mode. The system's observations, concerning performance and current actions (robot mode and alarm system status), and used for decision-making, are represented by the colors of the points on the curve.}
\end{figure*}
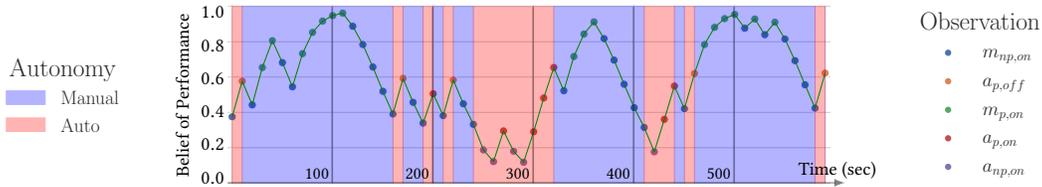

The experiments were conducted with four interaction control policies: robust POMDP solution, robust MDP solution, a random policy \cite{chanel2020towards} (the very same data collection policy), and a policy where the robot mode was fixed to automatic with alarms. These conditions were pseudo-randomized among participants. The robust POMDP policy was computed as detailed in Sec. \ref{sec:method} with $\gamma=\{0.7, 0.8, 0.9, 0.97, 0.98, 0.99\}$ and sampling $\mathrm{N}_M = 10000$ models. For each sampled model, each policy was deployed during $\mathrm{N}_{E} = 10000$ histories. The policy corresponding to $\gamma = 0.98$ was selected because it had the highest estimated Value-at-Risk at risk-level $0.5$. The MDP robust policy was obtained as follows: once the classifiers had been trained (third blue box in Fig. \ref{fig:pipeline}), instead of learning an observation function and an HMM dynamics, the classifiers were directly used to perform dimensionality reduction on the part of the data set split to learn the dynamics, converting hence the multi-dimensional and multi-modal time series of observations into time series of discrete states defining an MDP. On this transformed set of discrete MDP trajectories, several data-driven risk-sensitive baselines - e.g. soft-robust optimization \cite{lobo2021softrobust} - were executed, producing a set of candidate policies. Due to the low state-action space dimensionality, the EvC algorithm drove risk-sensitive policy selection, using the same criteria as for the POMDP case. Surprisingly, the robust MDP policy resulted in a fully manual mode with alarms, where the human operator was confronted to the multitasking paradigm during the whole mission. In contrast, with the Fixed Automatic (FA) mode interaction control policy, the human faces an easier task with respect to the other interaction conditions and is less prone to deleterious mental states. In practice, we would like to use the automatic mode only when the human operator is estimated unable to drive the robot, and not during the whole mission. Hence, although the results obtained with this policy are presented for discussion, its comparison with the policies calculated by our method is outside the scope of this work. After the four iterations, participants completed the KSS questionnaire again. The entire experiment lasted approximately 1 hour and 30 minutes per participant, totaling 39 hours.

\subsection{Results and Analysis}
\label{sec:discussion}
In this subsection, we elaborate on the results. The subsequent analysis will cover robustness, data informativeness, data scarcity, inter-subject variability, and the explainability of the implemented approach. We will not discuss the comparison with the FA policy because: (i) it was not learned from data using an ORL approach; (ii) 
it served to establish the best possible average score of the environment, in a focused sub-task distinct from other conditions;
(iii) in critical missions, such a policy is inadvisable due to the necessity of operator vigilance and control (for responsibility issues).

\paragraph{POMDP policy}
We start by analyzing qualitatively the POMDP risk-sensitive policy. With this policy, the controller changes the autonomy mode of the robot from manual mode to automatic mode when the estimated marginalized belief of performance $\beta_t$ falls below a given threshold - i.e. the system's belief is that the state is \textit{non-performant}. This policy, obtained with the paradigm previously outlined, is reasonable, since in $\mathcal{D}$, the time steps associated with the manual mode and performant operators were observed to be related to a higher score, while time steps with a non-performant operator had the highest reward when the robot was in automatic mode. For illustration purposes, an online $\beta_t$ computation during a mission run using the robust POMDP policy, along with observations received and deployed actions, is displayed in Figure \ref{fig:time_belief}.

\paragraph{Comparing the distribution of scores} Descriptive statistics for the results can be found in Table \ref{tab:descriptive_statistics}. We started by comparing the score distribution obtained with the random interaction policy evaluated in data set $\mathcal{D}_{val}$ with the data in the starting batch $\mathcal{D}$, that was also collected using a random policy (first line of Table \ref{tab:descriptive_statistics}). The Mann-Whitney U test was conducted to compare the two independent groups of scores. The test revealed a significant difference between the groups ($U = 686.0$, $p$-value $< 0.05$), rejecting the null hypothesis of no difference between the groups at the 5\% significance level. We can conclude that $\mathcal{D}$ and $\mathcal{D}_{val}$, even when the interacting policy is the same, do not follow the same distribution. This finding underscores the challenge of our task, as we had to work with a limited data set $\mathcal{D}$ and huge inter- and intra- subject variability.

\begin{table}[b]
\small
\centering
\caption{Descriptive statistics of the scores by interaction control policy (random or learned) and data set. $\mathcal{D}$ is the data set initially collected. $\mathcal{D}_{val}$ is the data set including the results of the experimental validation of the HITL-ORL pipeline. Best value per metric is displayed in \textbf{bold}. FA states for Fixed-Automatic control policy, leading to a single-task mission.}
\label{tab:descriptive_statistics}
\begin{tabular}{lrrrrrrr}
\toprule
{} &       mean &       std &   min & 25\% &   Median &    75\% &   max \\
Policy &                        &     &   &        &       &        &       \\
\midrule
Random ($\mathcal{D}$) &    22.1 &  10.0 &   1.0 & 12.8 &  26.5 &  29.3 &  36.0 \\
\midrule
Random ($\mathcal{D}_{val}$) &    17.9 &  9.6 &   4.0 & 8.3 &  20.0 &  26.8 &  32.0 \\
MDP ($\mathcal{D}_{val}$)   &    22.7 &  7.6 &   6.0 & 18.0 &  22.5 &  \textbf{28.8} &  \textbf{39.0} \\
POMDP ($\mathcal{D}_{val}$) &    \textbf{23.4} &  \textbf{5.7} &  \textbf{14.0} & \textbf{18.5} &  \textbf{24.0} &  27.5 &  37.0 \\
\midrule
FA ($\mathcal{D}_{val}$) & \textit{25.6} & \textit{5.0} & \textit{9.0} & \textit{25.3} & \textit{27.0} & \textit{28.0} & \textit{32.0} \\
\bottomrule
\end{tabular}
\end{table}

\paragraph{Robustness}
To assess whether applying an adaptive control policy leads to more robust results than deploying the policy used to collect the data, we performed a Friedman's statistical test to analyze the data of scores during the missions in $\mathcal{D}_{val}$. The results revealed a significant difference among the policies ($\chi^{2}(3)=13.07$, $p<0.01$). Post-hoc analyses using the Wilcoxon signed-rank test with a Hommel correction indicated that the Random policy (Median = $20$) yielded significantly lower scores compared to the MDP policy (Median = $22.5$, $p<0.05$), and the POMDP policy (Median = 24, $p<0.05$). Hence, it is convenient to deploy a policy learned with the proposed HITL-ORL method. Despite the better results of the POMDP policy for risk-sensitive metrics in Table \ref{tab:descriptive_statistics} (minimum score more than double the minimum one achieved with the MDP policy, and a lower standard deviation than the random/MDP policies), no statistically significant difference was observed between the MDP and the POMDP robust policies. The MDP policy maintains the manual mode, so the mission is therefore entirely multi-tasking, allowing users who are comfortable with this context to perform better, but making others perform very poorly. With the POMDP policy, the robot's mode depends on user state estimation (Fig. \ref{fig:time_belief}).

\begin{figure}[t]
\centering
\includegraphics[width=\linewidth]{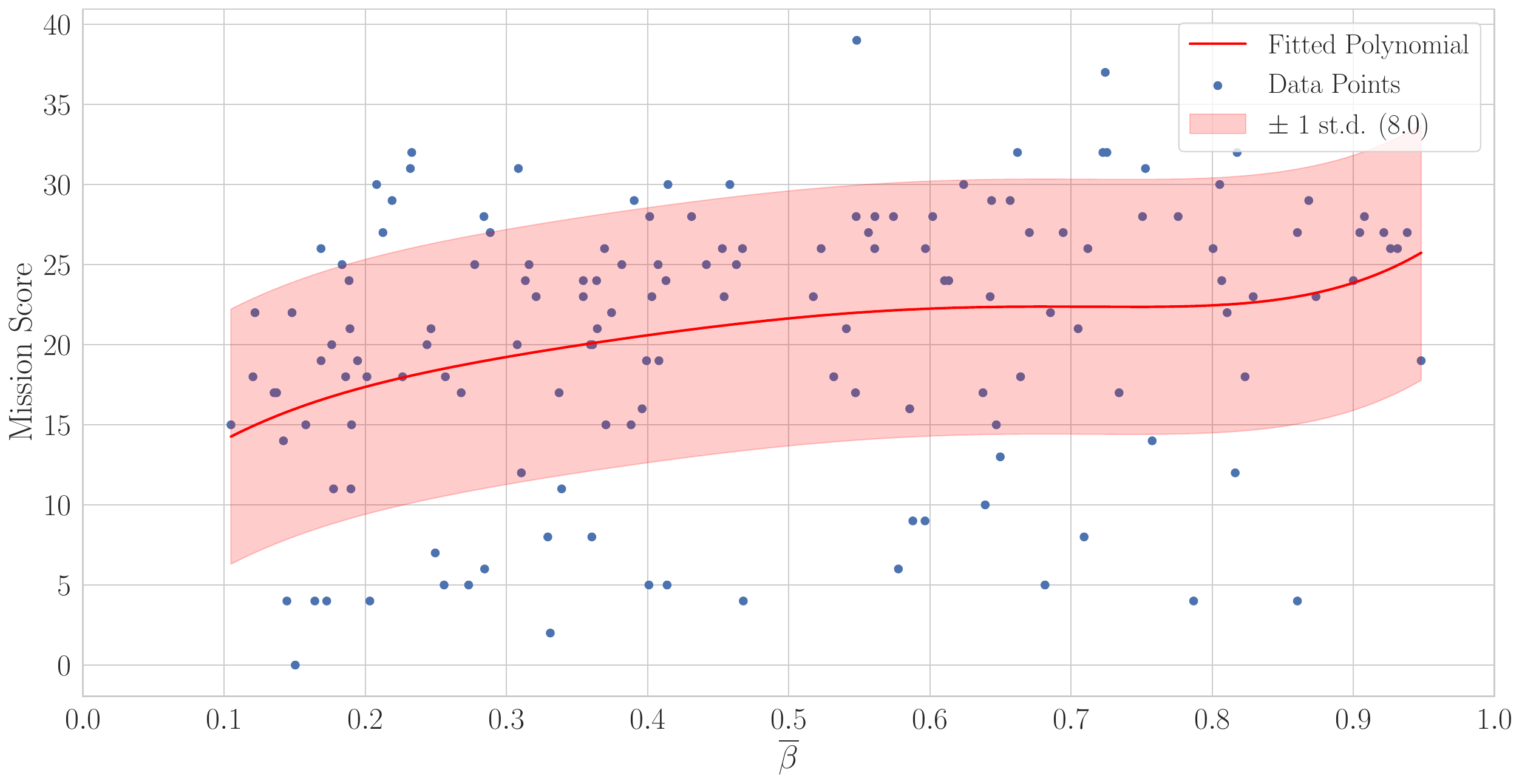}
\caption{Scatter plot displaying the relationship between the average belief of human-robot performance during a mission (x-axis) and the mission score (y-axis). A fifth-order (monotonic) polynomial is fitted to show the positive correlation between $\overline{\beta}$ and the score. The shaded area represents plus or minus one standard deviation of the residuals of the fit.}
\label{fig:corr_perf_score}
\Description{The x-axis represents the average belief in performance for a mission, and the y-axis represents the final score of that mission. The set of all missions is therefore represented by a scatter plot, and a polynomial curve is fitted to show the overall trend which is a positive correlation (increasing curve).}
\end{figure}

\paragraph{Validity of the POMDP models} To assess the POMDP models we calculate the POMDP belief for each time step, $\beta_t$, that represents the probability that during that time step the human-robot team is performant. The said $\beta_t$ is computed for each mission, including those run with a policy other than the POMDP one (obtained with EvC using SARSOP). If the model is accurate, we anticipate a high average belief of performance, $
\overline{\beta} = \frac{1}{T}\sum_{t=1}^{T}\beta_t$ 
in missions with high scores, and \textit{vice-versa}. Thus, we compute the correlation between $\overline{\beta}$ and the score for each mission. The Spearman's rank correlation reveals a significant positive correlation between $\overline{\beta}$ and the global mission score ($\rho(\overline{\beta}) = 0.325$, $N(\overline{\beta}) = 153$, $p(\overline{\beta})$-value $<0.001$). In Figure \ref{fig:corr_perf_score}, we illustrate the relationship between $\overline{\beta}$ and the mission score, along with a fitted (monotonic) fifth-order polynomial.
A positive correlation suggests alignment of the POMDP with actual events, despite classifiers with balanced accuracies around $0.67$. 
The Spearman's correlation coefficient only identifies monotonic correlations, without probing variable interdependence. To evaluate dependence between $\overline{\beta}$ and the score, we used the Randomized Dependence Coefficient (RDC) \cite{lopez2013randomized}, using suggested hyperparameters from the reference. The resulting RDC was $0.348$, almost identical to the Spearman's coefficient, suggesting a likely absence of non-monotonic dependence between the variables.

\paragraph{Analysis of the used classifiers and Explainability} The confusion matrices for the eight classifiers, four computed on the test set (in $\mathcal{D}$) and four on the data set obtained from the experimental results ($\mathcal{D}_{val}$) and split according to the same criterium were analyzed with a two-proportion Welch's t-test (refer to Supplementary Material). As the $p$-value is less than or equal to $0.05$ for almost all states, we can reject the null hypothesis that the classification outcomes have equivalent proportions. As a result, at the very least, the classifiers cannot be used for an \textit{accurate} representation of the process we aimed to model with both an MDP and/or a POMDP. This effect can be due to data scarcity and insufficient data informativeness, and also to an intense inter-subject variability both in the training set in $\mathcal{D}$ and in the set of experimental results $\mathcal{D}_{val}$.

\begin{figure}[b]
    \centering
        \includegraphics[width=\linewidth]{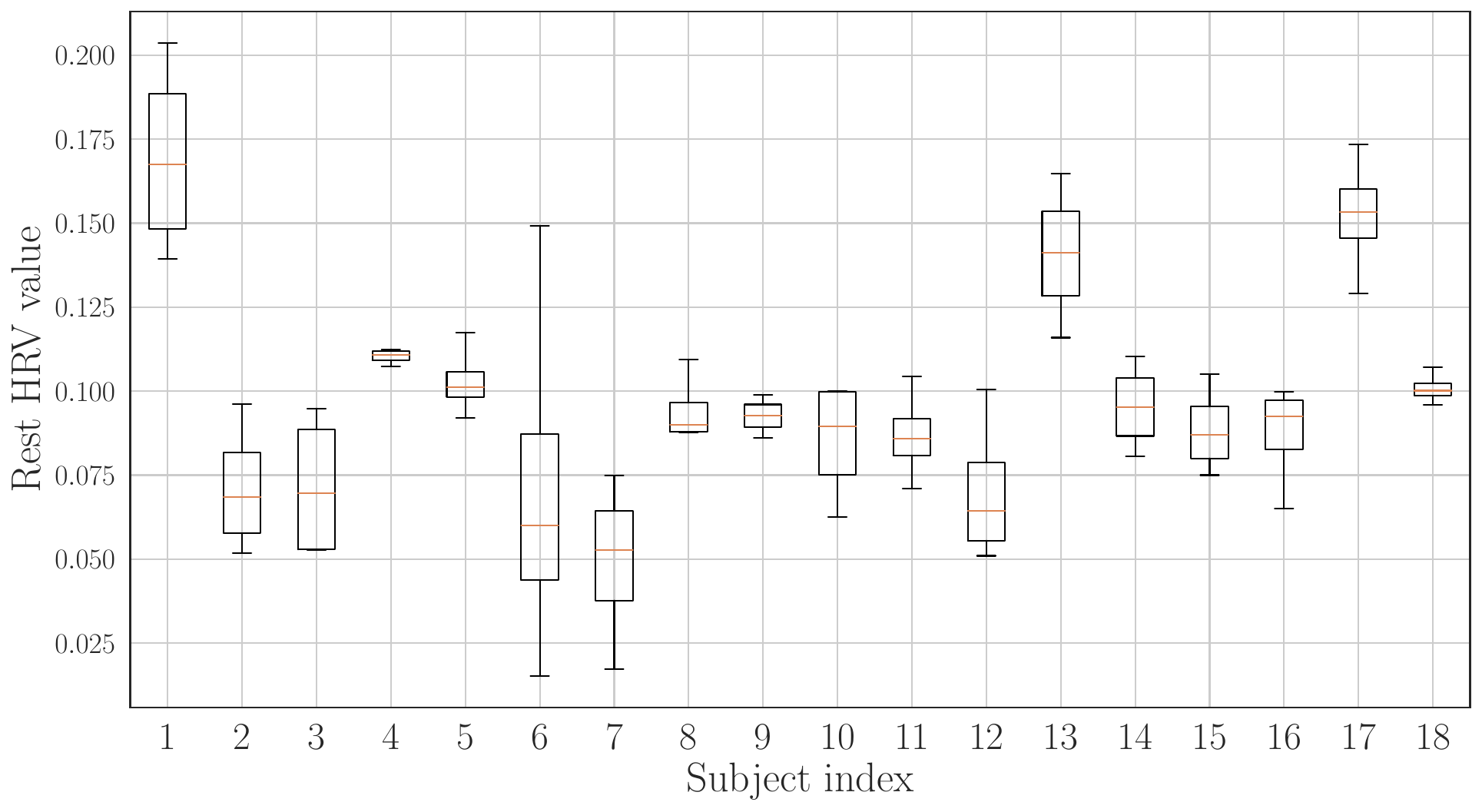}
        \caption{Rest HRV per subject before a mission. Since these values are used to normalize the features to be used by the ML model, variability could lead to a failure of the pipeline.}
        \label{fig:rest_hrv}
    \Description{The x-axis represents the participants, and the y-axis represents the HRV value during the resting session, just before the mission. This value is used to normalize the measurements of the following mission. Boxplots are used here to illustrate the variability of this HRV value.}
\end{figure}

To study which features were important for the classifier we used SHAP \cite{lundberg2020local} (see Supplementary Material).
The normalized HRV (a cardiac feature) in the 10-second time window is among the two most important features used for performance classification. Since the HRV is obtained after the standardization procedure of subtracting the resting baseline from the live measurement, any anomaly or perturbation in the standardization can have a huge impact on the classification. The explainability analysis provides great insights into the usefulness of incorporating physiological computing for HITL systems. However, inter-subject variability, and improper processing may lead to unwanted results. 
The inter-subject variability for the HRV at rest before each mission in the batch data set is displayed in Figure \ref{fig:rest_hrv}. In spite of the normalization attempts, we observed notable discrepancies in the recorded resting HR and HRV baselines for the same individual across different missions (see Supplementary Material).
The discrepancies were even more pronounced for HRV measurements at rest: the disparity between the maximum and minimum measurements of HRV at rest per subject could be as much as eightfold (see Subject 6 in Fig\ref{fig:rest_hrv}).

\paragraph{Subjective feedback from participants}
A Wilcoxon signed-rank test using the \textit{z-split} method was conducted to analyze the KSS scores before and after the experimental procedure. The test indicated that the difference between the two sets of scores was not statistically significant ($z=143.5$, $p=0.41$). This means that participants did not report subjective fatigue after the experiments, suggesting they remained committed to the task. For the NASA-TLX questionnaire scores, a Friedman's test was performed to analyze the overall rating, revealing a significant difference among the policies ($\chi^{2}(3)=36.98$, $p<0.001$). Post-hoc analysis using the Wilcoxon signed-rank test with Hommel correction showed a significant difference between the ratings obtained when the controller followed the fixed-automatic (FA) policy (Median $= 40.0$) and the other policies with $p < 0.001$ (Random, Median $= 61.2$; MDP, Median $= 71.0$; POMDP, Median $= 65.3$). These findings indicate that the FA policy, as expected, is perceived as the less demanding policy; and that the impact on the subjective workload of adaptive policies, unfortunately, remains inconclusive. Surprisingly, the results of the Fluency survey showed the FA policy (Median $=5.2$) was perceived as providing the most fluent interaction, while the MDP adaptive control policy is considered the least fluent (Median $=2.0$). A Friedman's test was conducted to analyze the overall rating, revealing a significant difference among the policies ($\chi^{2}(3)=32.72$, $p<0.001$). Post-hoc analysis using the Wilcoxon signed-rank test with Hommel correction showed a significant difference between the ratings obtained for every pair of policies ($p < 0.01$), except for the pair made of the Random (Median $= 4.0$) and the POMDP (Median $= 4.5$) policies ($p = 0.62$).
After discussion with participants, we speculate that FA policy (the easier experimental condition in which no adaptive interaction was used, as the robot was navigating autonomously all the time) favored some participants' experience. They reported that they could concentrate on the tank management tasks because they were confident in the robot's behavior, thus preferring such a constant task allocation policy. Despite this, we believe the findings on the random policy and adaptive POMDP policy comparison warrant further investigation to better understand the potential of the latter in enhancing human-robot interaction experience and performance (see Table \ref{tab:descriptive_statistics}). 

\section{Conclusions and Future Work}
\label{sec:conclusions}

In this work, we propose a methodology to approximate a POMDP model that includes a Bayesian representation of model uncertainty. We implement this through a specific use-case: the Firefighter Robot Game, a human-robot interaction proof-of-concept scenario. Physiological and behavioral features were integrated into the system, providing useful insights for calculating an interaction control policy, which determines the robot's autonomy during the mission. The control policies were obtained in an offline setting using only a previously collected set of human-robot interaction experiences. This data set comprised data from eighteen human operators who interacted with the system in laboratory facilities using a random interaction policy. The FRG is modeled as a POMDP, achieving an interpretable state space representation and a Bayesian representation of model uncertainty. 

Given the limited size of the data set and the relatively small number of volunteers, we assumed that we did not possess enough information to capture the full stochastic description of the environment. Therefore, we resorted to the Bayesian formalism to estimate uncertainty over the set of possible models, taking inspiration from the EvC \cite{angelotti2022an} method. More precisely, we extended the EvC policy selection method to consider model uncertainty in partially observable domains. To do this, we first computed a set of different policies for a data-driven trivial POMDP model using different discount factors ($\gamma$'s). Then, using Bayesian formalism and Monte Carlo sampling, we exploited model uncertainty to select a policy from among the set of candidate policies, according to a risk-sensitive metric.

We compared the robustness and the performance of the obtained adaptive POMDP policy with that used to collect the original data set, among others. This comparison was made by performing a new set of experiments involving twenty-six human operators in laboratory facilities. 
The experiments showcased that the proposed paradigm led to significantly higher performance than that obtained with the data collection policy. Furthermore, our results suggest that considering partial observability is propaedeutic to more robust policies. Moreover, by scrutinizing our experimental results, we identified several obstacles to the application of ORL, particularly highlighting the opportunities and challenges that arise when a human operator is included in the environment. Data representativeness and scarcity were major concerns; proper statistical tests were necessary to individuate these issues. A lack of sufficient diversity in data sets hindered generalization and would have required appropriate diverse recruitment campaigns and possibly more data. Nevertheless, the statistical tests showed a positive correlation between the belief in the human-robot team's performance and the mission score. 

Regarding the classifiers used to map high-dimensional
raw features into a low-dimensional observation space, we note that they relied heavily on physiological markers like standardized HRV for classification. The approach used to normalize HR and HRV features by subtracting resting baselines revealed potential discrepancies. As classifiers may favor these features, results could be compromised. On one hand, implementing an efficient experimental protocol that considers physiological and behavioral measurements and their variability among participants remains an open question. On the other hand, addressing partial human state observability and implementing ORL methods to handle model uncertainty can mitigate the effects of poor data curation and scarcity, thereby improving generalization.

\paragraph{Perspectives}
Future work could encompass modifying the FRG to allow the human operator to set the autonomy mode of the robot whenever desired, permitting thus to study the application of ORL and physiological computing to scenarios characterized by complete mixed-initiative human-robot interaction. Moreover, in the light of the current study, it might be appropriate to collect larger data sets before computing the risk-sensitive control policies with offline reinforcement learning, or to limit the analysis to a more homogeneous set of human operators in order to reduce inter-subject variability.

\begin{acks}
This work was funded by Artificial and Natural Intelligence Toulouse Institute (ANITI) - Institut 3iA (ANR-19-PI3A-0004) and financially supported by a chair from Dassault Aviation (CASAC).
\end{acks}

\bibliographystyle{ACM-Reference-Format} 
\bibliography{biblio}

\end{document}


\pagestyle{fancy}
\fancyhead{}

\maketitle 

\section{Features and state representation}
The raw features of the Firefighter Robot Game provided to the classifiers are the following, computed on 10-seconds time windows:
\begin{itemize}
\item \textit{nav}: the number of times the robot navigation keys have been typed by the human operator on the keyboard;
\item \textit{space}: the number of times the space key is pressed on the keyboard, used to throw water to put out fires;
\item \textit{HR}: the normalized Heart Rate;
\item \textit{HRV}: the normalized Heart Rate Variability;
\item \textit{nbAOI1-5} (5 features): the number of separate gaze fixations in the different Areas of Interest;
\item \textit{durAOI1-5} (5 features): the duration of separate gaze fixations in the different Areas of Interest;
\item \textit{trees}: the number of trees currently on fire;
\item \textit{tank}: the current amount of water in the water pool;
\item \textit{tank local score}: the current amount of water in the robot's water tank.
\end{itemize}
The system is modeled as a (PO)MDP with nine states $$\mathcal{S} = \{m_{np,off}, m_{np,on}, m_{p,off}, m_{np,on}, a_{np,off}, a_{np,on}, a_{p,off}, a_{p,on}, g\}$$ and four actions. $g$ is an artificial absorbing state indicating the gameover, introduced for modeling necessities.
\section{Analysis of the used classifiers and Explainability}
\subsection{Graphical representation of classifiers for dimensionality reduction}
\begin{figure}[!h]
\centering
\includegraphics[width=\columnwidth]{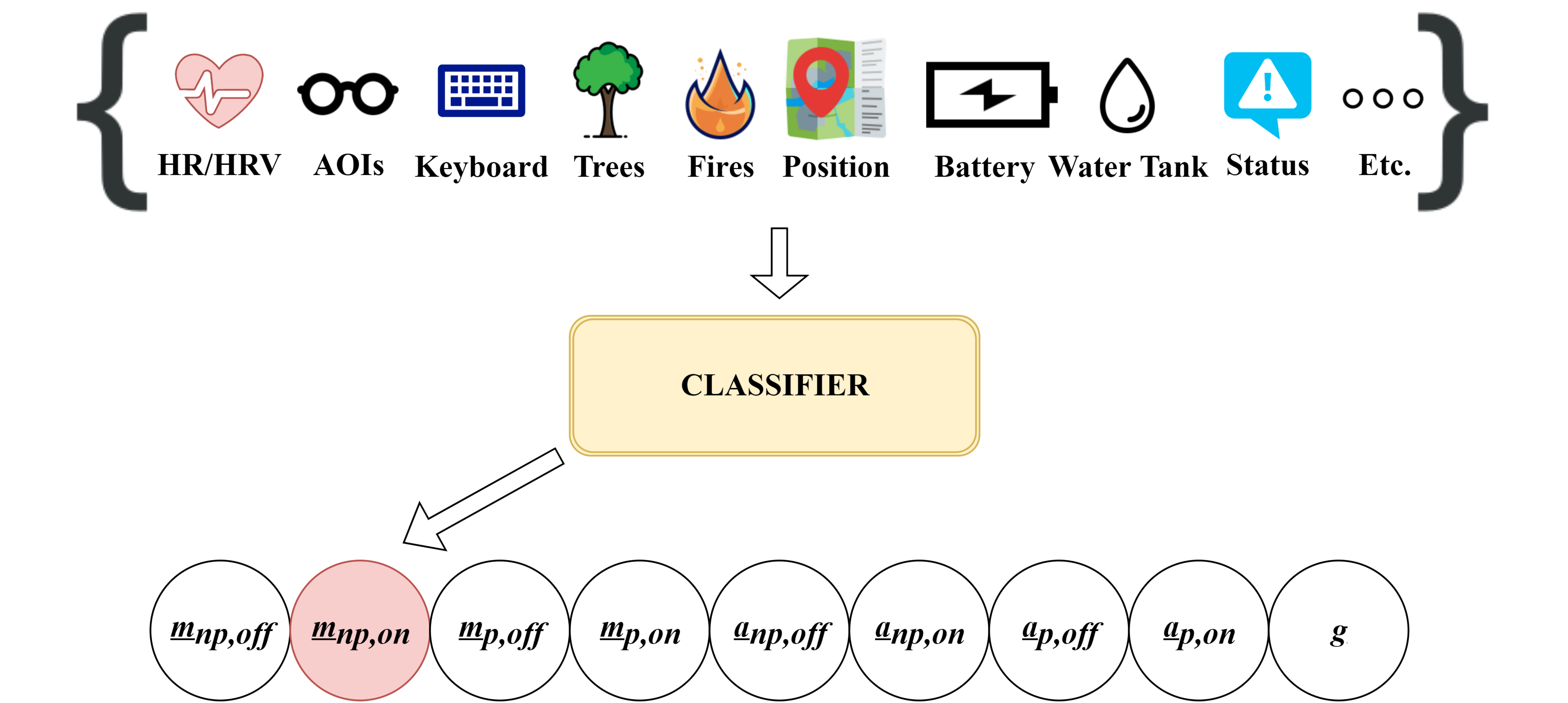}
\caption{Example of the procedure to map a high dimensional array of measurements regarding the human operator, the robot, the mission, and the interaction to one of the nine possible HMM/POMDP observations through a classifier. In the illustrated case, the classifier maps the high dimensional array of measurements (\textit{i.e.} the features $Y$ processed each 10-second time window) to the observation $\underline{m}_{np,on}$.}
\label{fig:classifier}
\end{figure}

In Figure \ref{fig:classifier} we provide a graphical representation of how the classifiers transform high-dimensional multimodal observations into low-dimensional discrete observations.
\FloatBarrier
\subsection{Confusion matrices of the classifiers}
In Table \ref{tab:obs_fun} we provide the confusion matrices of the four classifiers computed on the test set in $\mathcal{D}$ (see paragraph \textit{Classifier training} in Section 4).

\begin{table}[!h]
    \centering
    \caption{Confusion matrices computed on the test set for the different classifiers. }
    \label{tab:obs_fun}
    \begin{subtable}[b]{0.4\textwidth}
        \centering
                \caption{Manual, Alarms ON. Sensitivity: 0.60, Specificity: 0.72, Balanced Accuracy: 0.66.}
        \begin{tabular}{lcc}
            \toprule
                & Obs. Non-perf. & Obs. Perf. \\
            \midrule
            Non-perf. & 113 (0.72) & 43 (0.28) \\
            Perf.     & 71 (0.40) & 105 (0.60) \\
            \bottomrule
        \end{tabular}
    \end{subtable}
    \hfill
    \begin{subtable}[b]{0.4\textwidth}
        \centering
                \caption{Manual, Alarms OFF. Sensitivity: 0.71, Specificity: 0.67, Balanced Accuracy: 0.69.}
        \begin{tabular}{lcc}
            \toprule
                & Obs. Non-perf. & Obs. Perf. \\
            \midrule
            Non-perf. & 105 (0.67) & 51 (0.33) \\
            Perf.     & 51 (0.29) & 125 (0.71) \\
            \bottomrule
        \end{tabular}
    \end{subtable}

    \vspace{1cm}

    \begin{subtable}[b]{0.4\textwidth}
        \centering
                \caption{Auto, Alarms ON. Sensitivity: 0.75, Specificity: 0.59, Balanced Accuracy: 0.67.}
        \begin{tabular}{lcc}
            \toprule
                & Obs. Non-perf. & Obs. Perf. \\
            \midrule
            Non-perf. & 99 (0.59) & 70 (0.41) \\
            Perf.     & 46 (0.25) & 140 (0.75) \\
            \bottomrule
        \end{tabular}
    \end{subtable}
    \hfill
    \begin{subtable}[b]{0.4\textwidth}
        \centering
                \caption{Auto, Alarms OFF. Sensitivity: 0.64, Specificity: 0.70, Balanced Accuracy: 0.67.}
        \begin{tabular}{lcc}
            \toprule
                & Obs. Non-perf. & Obs. Perf. \\
            \midrule
            Non-perf. & 91 (0.70) & 39 (0.30) \\
            Perf.     & 55 (0.36) & 97 (0.64) \\
            \bottomrule
        \end{tabular}
    \end{subtable}
\end{table}

In Table \ref{tab:res_conf_mat} we provide the confusion matrices of the four classifiers computed on a test set in $\mathcal{D}_{val}$ obtained applying the same splitting criteria (see paragraph \textit{Data split} in Section 4).

\begin{table}[!h]
    \centering
    \caption{Confusion matrices computed on the experimental results for the different classifiers. In parentheses the row normalized values.}
    \label{tab:res_conf_mat}
    \begin{subtable}[b]{0.4\textwidth}
        \centering
        \caption{Manual mode, Alarms ON. Sensitivity: 0.38, Specificity: 0.81, Balanced Accuracy: 0.60.}
        \begin{tabular}{lcc}
            \toprule
                & Obs. Non-perf. & Obs. Perf. \\
            \midrule
            Non-perf. & 1484 (0.81) & 346 (0.19) \\
            Perf.     & 349 (0.62) & 217 (0.38) \\
            \bottomrule
        \end{tabular}
    \end{subtable}
    \hskip 40pt
    \begin{subtable}[b]{0.4\textwidth}
        \centering
        \caption{Manual mode, Alarms OFF. Sensitivity: 0.37, Specificity: 0.69, Balanced Accuracy: 0.53.}
        \begin{tabular}{lcc}
            \toprule
                & Obs. Non-perf. & Obs. Perf. \\
            \midrule
            Non-perf. & 108 (0.69) & 49 (0.31) \\
            Perf.     & 43 (0.63) & 25 (0.37) \\
            \bottomrule
        \end{tabular}
    \end{subtable}

    \vspace{1cm}

    \begin{subtable}[b]{0.4\textwidth}
        \centering
        \caption{Auto mode, Alarms ON. Sensitivity: 0.54, Specificity: 0.50, Balanced Accuracy: 0.52.}
        \begin{tabular}{lcc}
            \toprule
                & Obs. Non-perf. & Obs. Perf. \\
            \midrule
            Non-perf. & 383 (0.50) & 376 (0.50) \\
            Perf.     & 146 (0.46) & 168 (0.54) \\
            \bottomrule
        \end{tabular}
    \end{subtable}
        \hskip 40pt
    \begin{subtable}[b]{0.4\textwidth}
        \centering
        \caption{Auto mode, Alarms OFF. Sensitivity: 1.00, Specificity: 0.01, Balanced Accuracy: 0.50.}
        \begin{tabular}{lcc}
            \toprule
                & Obs. Non-perf. & Obs. Perf. \\
            \midrule
            Non-perf. & 1 (0.01) & 167 (0.99) \\
            Perf.     & 0 (0.00) & 70 (1.00) \\
            \bottomrule
        \end{tabular}
    \end{subtable}
\end{table}
\subsection{Two-proportion Welch's t-test}
\begin{table}[h]
    \caption{Results of the two-proportion Welch's t-test for the eight states using the datasets $\mathcal{D}$ and $\mathcal{D}_{val}$. In the states, $m$ or $a$ stands for the manual or automatic autonomy level of the robot, $p$ or $np$ stands for performant or not performant participant, and $on$ or $off$ stands for the alarm notification status. Bold values indicate a $p$-value less than 0.05.}
    \label{tab:welch_results}
    
        \centering
        \begin{tabular}{lccc}
            \toprule
            State & t-test ($t$) & df & $p$-value \\
            \midrule
            $\boldsymbol{m_{np, on}}$ & $\boldsymbol{-2.34}$ & $\boldsymbol{662.90}$ & $\boldsymbol{0.02}$ \\
            $\boldsymbol{m_{p, on}}$ & $\boldsymbol{5.05}$ & $\boldsymbol{804.41}$ & $\boldsymbol{<0.001}$ \\
            $m_{np, off}$ & $-0.28$ & $268.13$ & $0.78$ \\
            $\boldsymbol{m_{p, off}}$ & $\boldsymbol{5.06}$ & $\boldsymbol{122.87}$ & $\boldsymbol{<0.001}$ \\
            $\boldsymbol{a_{np, on}}$ & $\boldsymbol{1.93}$ & $\boldsymbol{478.27}$ & $\boldsymbol{0.05}$ \\
            $\boldsymbol{a_{p, on}}$ & $\boldsymbol{5.14}$ & $\boldsymbol{585.54}$ & $\boldsymbol{<0.001}$ \\
            $\boldsymbol{a_{np, off}}$ & $\boldsymbol{17.08}$ & $\boldsymbol{485.54}$ & $\boldsymbol{<0.001}$ \\
            $\boldsymbol{a_{p, off}}$ & $\boldsymbol{-9.28}$ & $\boldsymbol{167.00}$ & $\boldsymbol{<0.001}$ \\
            \bottomrule
        \end{tabular}
    
\end{table}
In Table \ref{tab:welch_results} we provide the results of the two-proportion Welch's t-test for the confusion matrices obtained on the test set in $\mathcal{D}$ and a test set in $\mathcal{D}_{val}$ obtained applying the same splitting criteria. 
As the $p$-value is less than or equal to $0.05$ for all states except for $m_{np, off}$, which is never encountered during a mission using adaptive policies, we can reject the null hypothesis that the classification outcomes have equivalent proportions. As a result, at the very least, the \textit{trivial} observation functions cannot be considered an accurate representation of the process we aimed to model with a POMDP. However, anticipating these findings is precisely why we developed the risk-sensitive pipeline.
\FloatBarrier
\subsection{SHAP Explainability analysis of features used by the classifiers}
\begin{figure}[!ht]
\centering
     \begin{subfigure}[t]{0.48\textwidth}
         \resizebox {\textwidth} {!} {
         \input{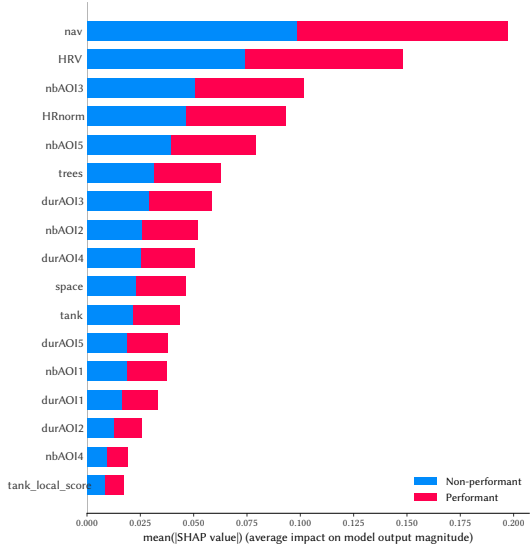}
         }
         \caption{Manual Mode - Alarms Off}
         \label{fig:shap_moff}
     \end{subfigure}
     \hskip 10pt
     \begin{subfigure}[t]{0.48\textwidth}
         \resizebox {\textwidth} {!} {
         \input{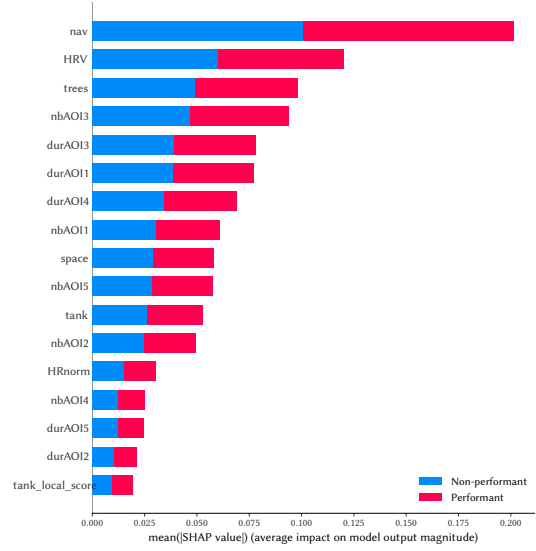}
         }
         \caption{Manual Mode - Alarms On}
         \label{fig:shap_mon}
     \end{subfigure}
     \\
     \begin{subfigure}[t]{0.48\textwidth}
         \resizebox {\textwidth} {!} {
         \input{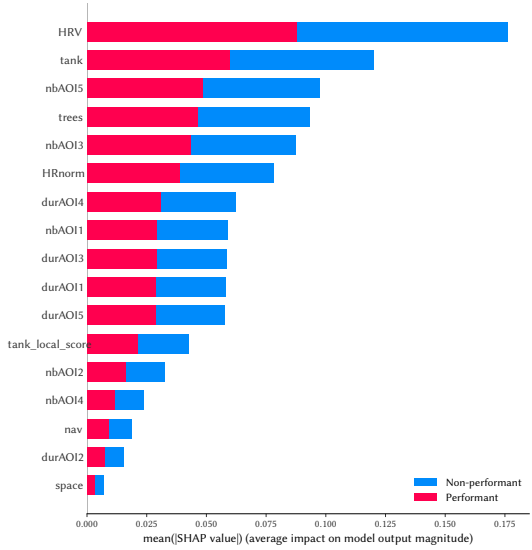}
	}
	\caption{Automatic Mode - Alarms Off}
         \label{fig:shap_aoff}
	\end{subfigure}
	\hskip 10pt
     \begin{subfigure}[t]{0.48\textwidth}
         \resizebox {\textwidth} {!} {
         \input{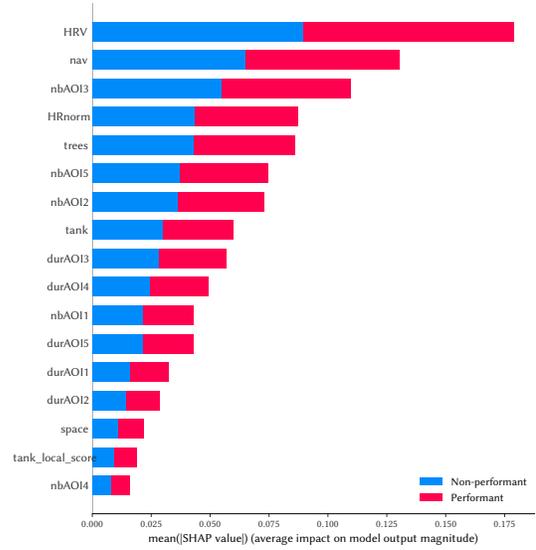}
}
		\caption{Automatic Mode - Alarms On}
         \label{fig:shap_aon}
     \end{subfigure}
	\caption{Shapley values for the features of each classifier (autonomy mode/notification status) obtained via SHAP's Tree Explainer. The features include \textit{HRV} (Normalized Heart Rate Variability), \textit{HRnorm} (Normalized Heart Rate), \textit{durAOI$_{i}$} (duration of fixations in AOI $i$), \textit{nbAOI}$_{i}$ (number of fixations in AOI $i$), \textit{nav} (number of times a navigation key has been pressed), \textit{space} (number of times the space bar has been pressed), \textit{trees} (number of extinguished fires), \textit{tank} (level of the water tank), and \textit{tank\textunderscore local\textunderscore score} (changes in the level of the water tank).}
	\label{fig:shap_rff}
\end{figure}

In Figure \ref{fig:shap_rff} we display the results of SHAP analysis of features used by the classifiers for classification (\text{i.e.} dimensionality reduction).
\FloatBarrier
\section{Trivial POMDP Model}
\subsection{Transition Function}
In Tables \ref{tab:action_manual_on}-\ref{tab:action_auto_off} we show the transition probabilities of the POMDP learned from the trivial observation functions.

\begin{table}[!h]
\centering
\caption{Transition probabilities for action: put manual mode and activate alarms. Rows are current states and columns are next states.}
\label{tab:action_manual_on}
\begin{tabular}{c|c|c|c|c|c|c|c|c|c}
 & $m_{np,off}$ & $m_{np,on}$ & $m_{p,off}$ & $m_{p,on}$ & $a_{np,off}$ & $a_{np,on}$ & $a_{p,off}$ & $a_{p,on}$ & $g$ \\
\hline
$m_{np,off}$ & 0 & 0.917 & 0 & 0.067 & 0 & 0 & 0 & 0 & 0.017 \\
$m_{np,on}$  & 0 & 0.917 & 0 & 0.065 & 0 & 0 & 0 & 0 & 0.019 \\
$m_{p,off}$  & 0 & 0.027 & 0 & 0.956 & 0 & 0 & 0 & 0 & 0.017 \\
$m_{p,on}$   & 0 & 0.032 & 0 & 0.956 & 0 & 0 & 0 & 0 & 0.012 \\
$a_{np,off}$ & 0 & 0.945 & 0 & 0.051 & 0 & 0 & 0 & 0 & 0.005 \\
$a_{np,on}$  & 0 & 0.931 & 0 & 0.063 & 0 & 0 & 0 & 0 & 0.006 \\
$a_{p,off}$  & 0 & 0.018 & 0 & 0.973 & 0 & 0 & 0 & 0 & 0.009 \\
$a_{p,on}$   & 0 & 0.016 & 0 & 0.965 & 0 & 0 & 0 & 0 & 0.020 \\
$g$          & 0 & 0 & 0 & 0 & 0 & 0 & 0 & 0 & 1 \\
\hline
\end{tabular}
\end{table}
\begin{table}[!h]
\centering
\caption{Transition probabilities for action: put manual mode and deactivate alarms. Rows are current states and columns are next states.}
\label{tab:action_manual_off}
\begin{tabular}{c|c|c|c|c|c|c|c|c|c}
 & $m_{np,off}$ & $m_{np,on}$ & $m_{p,off}$ & $m_{p,on}$ & $a_{np,off}$ & $a_{np,on}$ & $a_{p,off}$ & $a_{p,on}$ & $g$ \\
\hline
$m_{np,off}$ & 0.914 & 0 & 0.069 & 0 & 0 & 0 & 0 & 0 & 0.017 \\
$m_{np,on}$  & 0.914 & 0 & 0.067 & 0 & 0 & 0 & 0 & 0 & 0.018 \\
$m_{p,off}$  & 0.020 & 0 & 0.963 & 0 & 0 & 0 & 0 & 0 & 0.017 \\
$m_{p,on}$   & 0.024 & 0 & 0.964 & 0 & 0 & 0 & 0 & 0 & 0.012 \\
$a_{np,off}$ & 0.940 & 0 & 0.055 & 0 & 0 & 0 & 0 & 0 & 0.005 \\
$a_{np,on}$  & 0.923 & 0 & 0.071 & 0 & 0 & 0 & 0 & 0 & 0.006 \\
$a_{p,off}$  & 0.015 & 0 & 0.976 & 0 & 0 & 0 & 0 & 0 & 0.009 \\
$a_{p,on}$   & 0.013 & 0 & 0.967 & 0 & 0 & 0 & 0 & 0 & 0.020 \\
$g$          & 0 & 0 & 0 & 0 & 0 & 0 & 0 & 0 & 1 \\
\hline
\end{tabular}
\end{table}
\begin{table}[!h]
\centering
\caption{Transition probabilities for action: put automatic mode and activate alarms. Rows are current states and columns are next states.}
\label{tab:action_auto_on}
\begin{tabular}{c|c|c|c|c|c|c|c|c|c}
 & $m_{np,off}$ & $m_{np,on}$ & $m_{p,off}$ & $m_{p,on}$ & $a_{np,off}$ & $a_{np,on}$ & $a_{p,off}$ & $a_{p,on}$ & $g$ \\
\hline
$m_{np,off}$ & 0 & 0 & 0 & 0 & 0.941 & 0 & 0.042 & 0 & 0.017 \\
$m_{np,on}$  & 0 & 0 & 0 & 0 & 0.940 & 0 & 0.042 & 0 & 0.018 \\
$m_{p,off}$  & 0 & 0 & 0 & 0 & 0.020 & 0 & 0.963 & 0 & 0.017 \\
$m_{p,on}$   & 0 & 0 & 0 & 0 & 0.025 & 0 & 0.963 & 0 & 0.012 \\
$a_{np,off}$ & 0 & 0 & 0 & 0 & 0.921 & 0 & 0.074 & 0 & 0.005 \\
$a_{np,on}$  & 0 & 0 & 0 & 0 & 0.916 & 0 & 0.079 & 0 & 0.006 \\
$a_{p,off}$  & 0 & 0 & 0 & 0 & 0.023 & 0 & 0.968 & 0 & 0.009 \\
$a_{p,on}$   & 0 & 0 & 0 & 0 & 0.019 & 0 & 0.961 & 0 & 0.020 \\
$g$          & 0 & 0 & 0 & 0 & 0 & 0 & 0 & 0 & 1 \\
\hline
\end{tabular}
\end{table}
\begin{table}[!h]
\centering
\caption{Transition probabilities for action: put automatic mode and deactivate alarms. Rows are current states and columns are next states.}
\label{tab:action_auto_off}
\begin{tabular}{c|c|c|c|c|c|c|c|c|c}
 & $m_{np,off}$ & $m_{np,on}$ & $m_{p,off}$ & $m_{p,on}$ & $a_{np,off}$ & $a_{np,on}$ & $a_{p,off}$ & $a_{p,on}$ & $g$ \\
\hline
$m_{np,off}$ & 0 & 0 & 0 & 0 & 0.944 & 0 & 0.039 & 0 & 0.017 \\
$m_{np,on}$  & 0 & 0 & 0 & 0 & 0.943 & 0 & 0.038 & 0 & 0.018 \\
$m_{p,off}$  & 0 & 0 & 0 & 0 & 0.021 & 0 & 0.963 & 0 & 0.017 \\
$m_{p,on}$   & 0 & 0 & 0 & 0 & 0.025 & 0 & 0.962 & 0 & 0.012 \\
$a_{np,off}$ & 0 & 0 & 0 & 0 & 0.932 & 0 & 0.064 & 0 & 0.005 \\
$a_{np,on}$  & 0 & 0 & 0 & 0 & 0.929 & 0 & 0.066 & 0 & 0.006 \\
$a_{p,off}$  & 0 & 0 & 0 & 0 & 0.031 & 0 & 0.961 & 0 & 0.009 \\
$a_{p,on}$   & 0 & 0 & 0 & 0 & 0.026 & 0 & 0.954 & 0 & 0.020 \\
$g$          & 0 & 0 & 0 & 0 & 0 & 0 & 0 & 0 & 1 \\
\hline
\end{tabular}
\end{table}
\subsection{Reward Function}
In Table \ref{tab:rew} we report the average number of extinguished fires per modality and per time step observed in $\mathcal{D}$ that was used to define the reward function of the POMDP.
\begin{table}[h]
\centering
\caption{POMDP rewards for each hidden state, independently of the action.}
\label{tab:rew}
\begin{tabular}{|l|ccccccccc|}
\hline
\textbf{State} & $m_{np,off}$ & $m_{np,on}$ & $m_{p,off}$ & $m_{p,on}$ & $a_{np,off}$ & $a_{np,on}$ & $a_{p,off}$ & $a_{p,on}$  & $g$ \\
\hline
\textbf{Reward} &  0.257 & 0.289 & 0.603 & 0.741 & 0.257 & 0.333 & 0.432 & 0.493 & 0.0 \\
\hline
\end{tabular}
\end{table}
\FloatBarrier
\section{POMDP Solving and Policy Selection}
\subsection{Offline Policy Evaluation}
In Table \ref{tab:rff_policy_comparison} we show the comparison between different policies tried in simulation on the models sampled abiding by the procedure described in the main paper.
\begin{table}[ht]
\centering
\caption{Comparison of different policies ran in simulation. SARSOP means that the policy is the one obtained by solving the trivial POMDP with SARSOP and the indicated $\gamma$. Different robust and performance metrics are computed on the empirical distribution of returns. The best values in each column are in \textbf{bold}.}
\label{tab:rff_policy_comparison}
\begin{tabular}{lcccccccc}
\hline
Policy & Mean & Std. Dev. & Min & 25\% & Median & 75\% & Max \\
\hline
Random & 19.6 & \textbf{10.7} & \textbf{0.3} & 9.9 & 21.7 & 29.0 & 38.3 \\
SARSOP $\gamma = 0.70$ & 23.3 & 13.8 & \textbf{0.3} & 10.4 & 25.0 & \textbf{35.4} & \textbf{44.5} \\
SARSOP $\gamma = 0.80$ & 23.4 & 13.8 & \textbf{0.3} & 10.5 & 25.1 & \textbf{35.4} & \textbf{44.5} \\
SARSOP $\gamma = 0.90$ & 23.6 & 13.6 & \textbf{0.3} & 11.1 & 25.5 & \textbf{35.4} & \textbf{44.5} \\
SARSOP $\gamma = 0.97$ & 23.7 & 13.3 & \textbf{0.3} & 11.8 & 26.0 & 35.2 & \textbf{44.5} \\
SARSOP $\gamma = 0.98$ & \textbf{23.8} & 13.1 & \textbf{0.3} & 12.0 & \textbf{26.1} & 34.5 & \textbf{44.5} \\
SARSOP $\gamma = 0.99$ & 23.3 & 12.0 & \textbf{0.3} & \textbf{13.6} & 25.1 & 33.0 & \textbf{44.5} \\
SARSOP $\gamma = 0.999$ & 21.4 & 10.8 & \textbf{0.3} & \textbf{13.6} & 22.8 & 29.9 & \textbf{44.5} \\
\hline
\end{tabular}
\end{table}
\subsection{Robust POMDP policy}
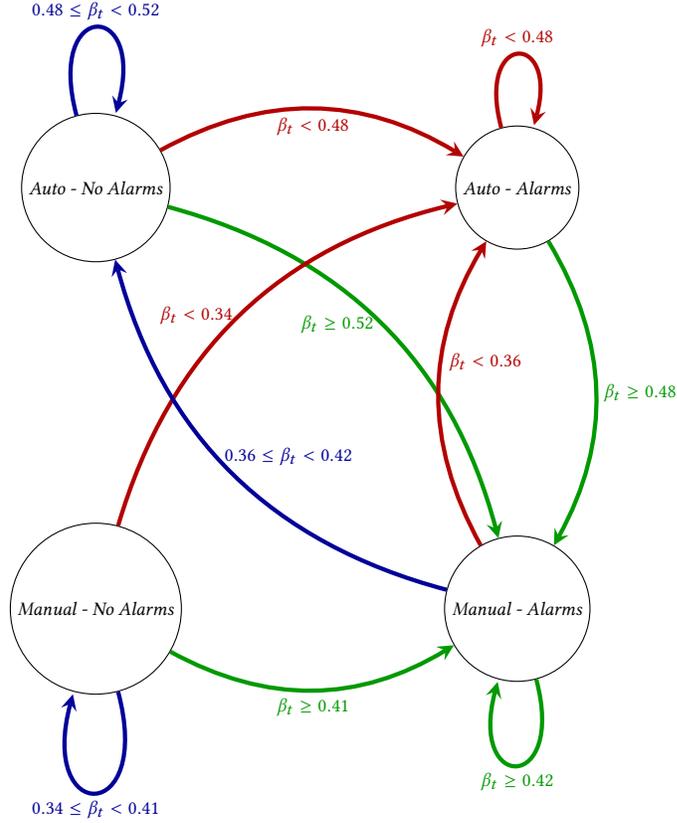
\begin{figure}[!h]
    \begin{center}
    \begin{tikzpicture}[>=stealth,node distance=7cm,on grid,auto, scale=0.8, transform shape]
  \node[state] (auto_no) {\textit{Auto - No Alarms}};
  \node[state] (auto_al) [right=of auto_no] {\textit{Auto - Alarms}};
  \node[state] (manual_no) [below=of auto_no] {\textit{Manual - No Alarms}};
  \node[state] (manual_al) [right=of manual_no] {\textit{Manual - Alarms}};
  \path[->, ultra thick]
    (auto_no) edge[bend left=30, green!60!black] node[pos=0.5, left, green!60!black, font=\bfseries] {$\beta_t \geq 0.52$} (manual_al)
              edge[loop above, blue!60!black] node[pos=0.5, blue!60!black, font=\bfseries] {$0.48 \leq \beta_t < 0.52$} (auto_no)
              edge[bend left=30, red!70!black] node[pos=0.5, below, red!70!black, font=\bfseries] {$\beta_t < 0.48$} (auto_al)
    (auto_al) edge[bend left=30, green!60!black] node[pos=0.5, right, green!60!black, font=\bfseries] {$\beta_t \geq 0.48$} (manual_al)
              edge[loop above, red!70!black] node[pos=0.5, red!70!black, font=\bfseries] {$\beta_t < 0.48$} (auto_al)
    (manual_no) edge[bend right=30, green!60!black] node[pos=0.5, below, green!60!black, font=\bfseries] {$\beta_t \geq 0.41$} (manual_al)
                edge[loop below, blue!60!black] node[pos=0.5, blue!60!black, font=\bfseries] {$0.34 \leq \beta_t < 0.41$} (auto_no)
                edge[bend left=30, red!70!black] node[pos=0.5, left, red!70!black, font=\bfseries] {$\beta_t < 0.34$} (auto_al)
    (manual_al) edge[loop below, green!60!black] node[pos=0.5, green!60!black, font=\bfseries] {$\beta_t \geq 0.42$} (manual_al)
                edge[bend left=30, blue!60!black] node[pos=0.55, right, blue!60!black, font=\bfseries] {$0.36 \leq \beta_t < 0.42$} (auto_no)
                edge[bend left=30, red!70!black] node[pos=0.6, right, red!70!black, font=\bfseries] {$\beta_t < 0.36$} (auto_al);
\end{tikzpicture}
    \end{center}
    \caption{Selected robust POMDP policy ($\gamma = 0.98$). Which action is taken and when? Let $\beta_t$ be the belief of performance for an observed state (Manual/Auto - Alarms/No Alarms) and time step $t$. Different values of $\beta_t$ lead to transitions to different autonomy and alarm modes. It is worth noting that \textit{Auto - Alarms} is an  absorbing modality for non performant missions, while \textit{Manual - Alarms} is an absorbing modality for performant missions. Moreover, since no arrows lead to \textit{Manual - No Alarms} and the initial modality \textit{Manual - Alarms}, the former modality is never reached during a mission that follows the robust POMDP policy.}
    \label{fig:pomdp-policy}
\end{figure}

In Figure \ref{fig:pomdp-policy} we show the policy obtained by solving the trivial POMDP with $\gamma = 0.98$ that was selected among the set of candidate ones according to the risk-sensitive criteria (best VaR${}_{0.5}$).
\FloatBarrier
\section{Physiological Measurements}
\subsection{HR at rest Baseline}
\begin{figure}[htbp]
        \centering
        \includegraphics[width=\linewidth]{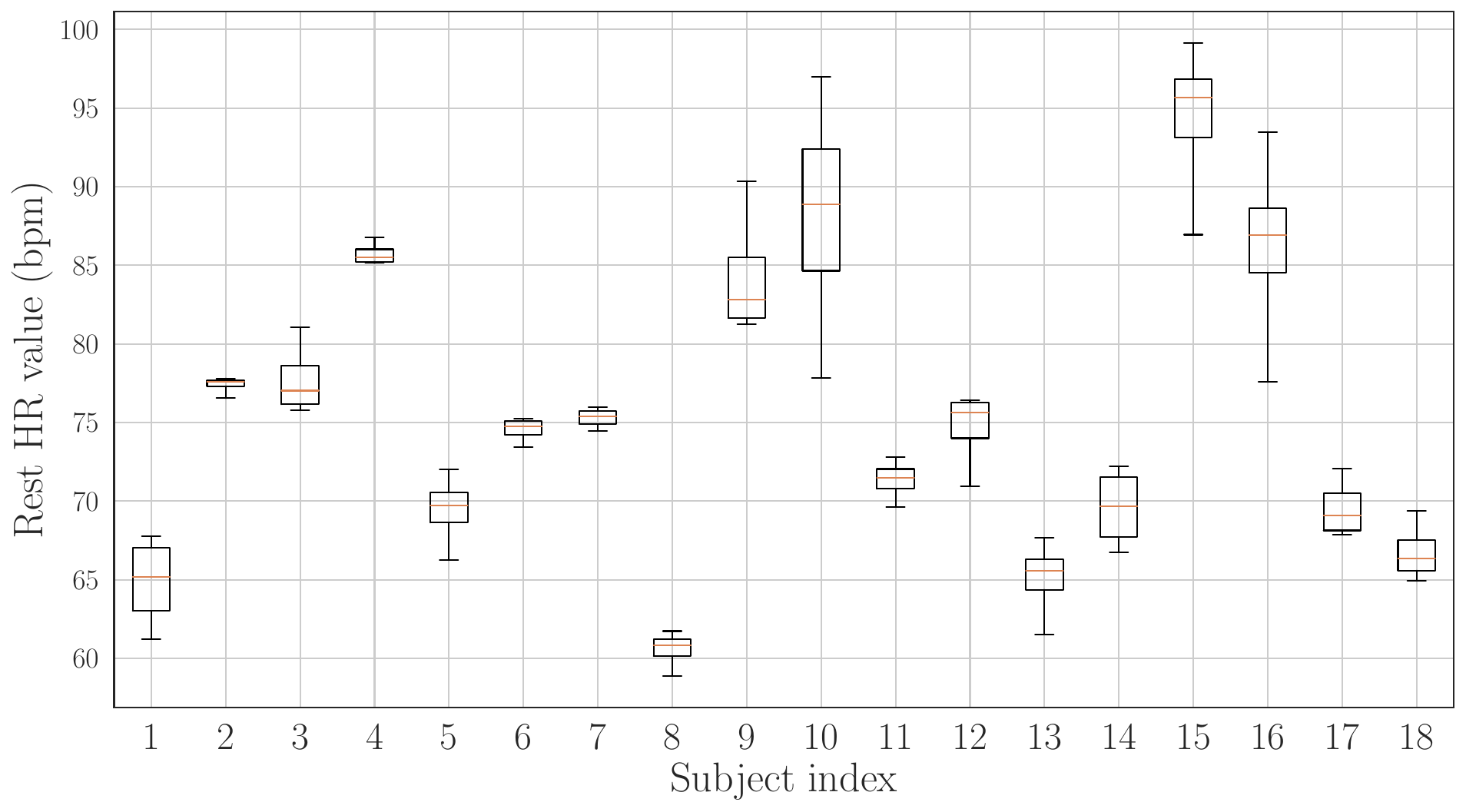}
        \caption{Rest average Heart Rate (bpm) per subject before a mission.}
        \label{fig:rest_hr}
\end{figure}

The HR at rest for subject before mission is displayed in Figure \ref{fig:rest_hr}.
\subsection{Eye-tracker}
\begin{figure}
    \centering

    \begin{subfigure}{0.85\linewidth}
        \includegraphics[trim={0 0.5cm 0 0},clip,width=\linewidth]{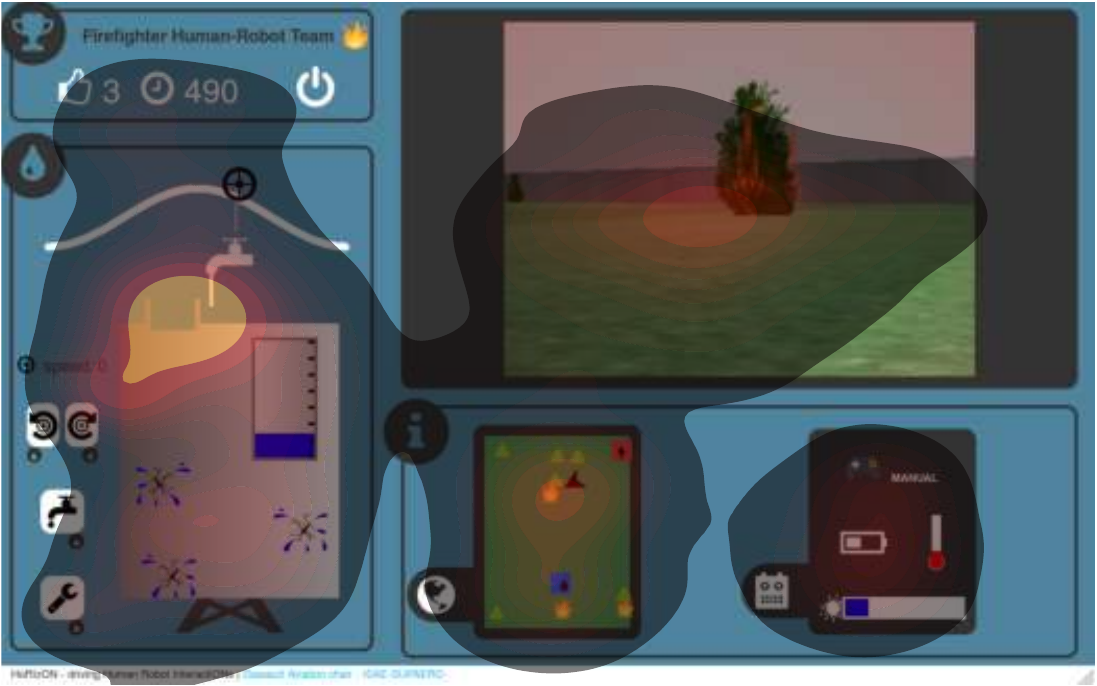}
    \end{subfigure}
    \begin{subfigure}{0.1\linewidth}
        \includegraphics[width=\linewidth]{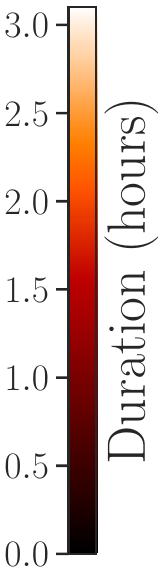}

    \end{subfigure}%

    \caption{Aggregate density of fixations on the monitor for missions using the adaptive POMDP policy.}
    \label{fig:eyetracker}
\end{figure}

Aggregate density of fixations on the monitor for missions using the adaptive POMDP policy is displayed in Figure \ref{fig:eyetracker}.
\FloatBarrier
\section{Subjective Feedbacks}
\begin{figure}[h]
\centering
\includegraphics[width=\columnwidth]{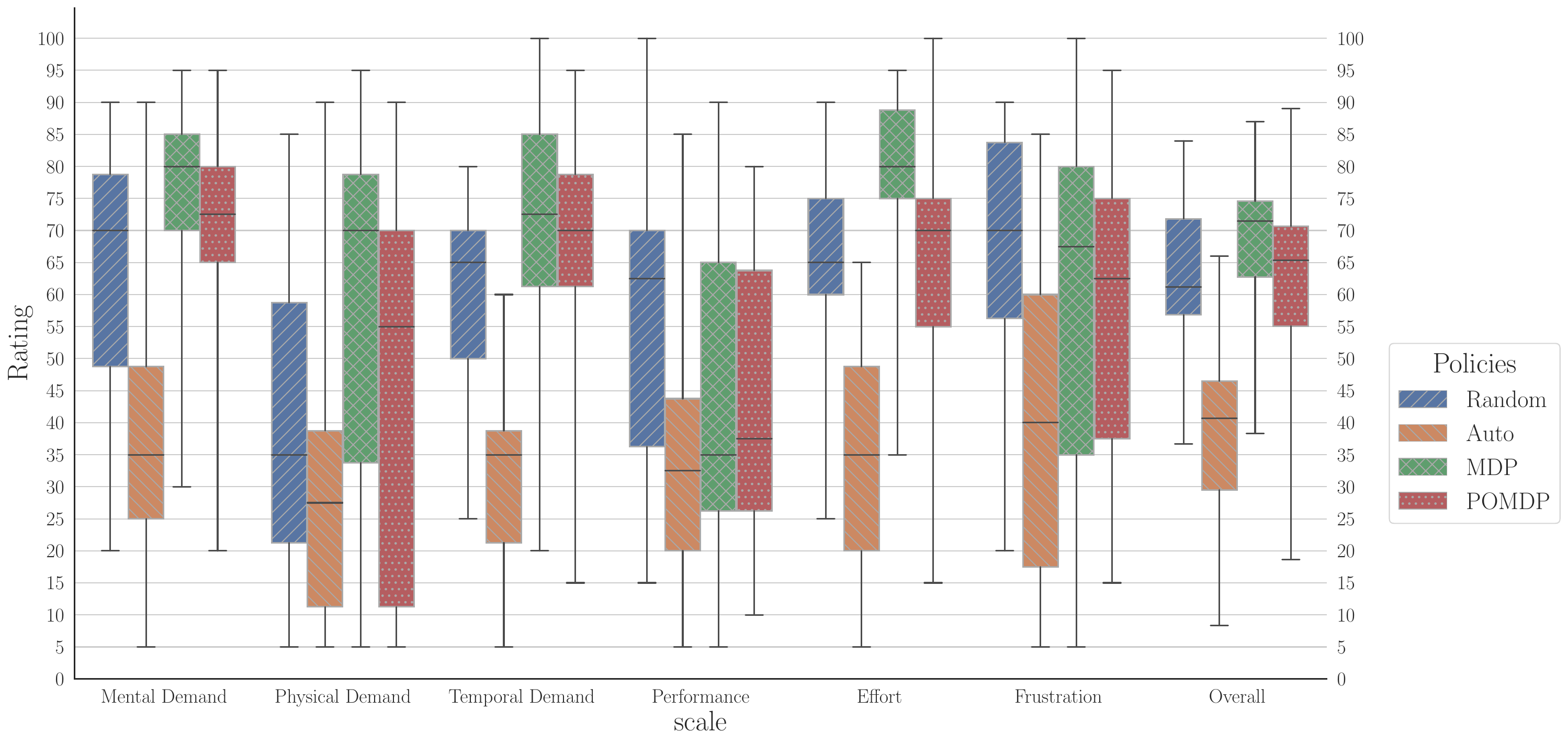}
\caption{Boxplot with the distributions of NASA-TLX by interaction control policy and scale. The Fixed Automatic policy (FA or Auto) is the least demanding across all scales.}
\label{fig:nasatlx}
\end{figure}

\begin{figure}[h]
\centering
\includegraphics[width=\columnwidth]{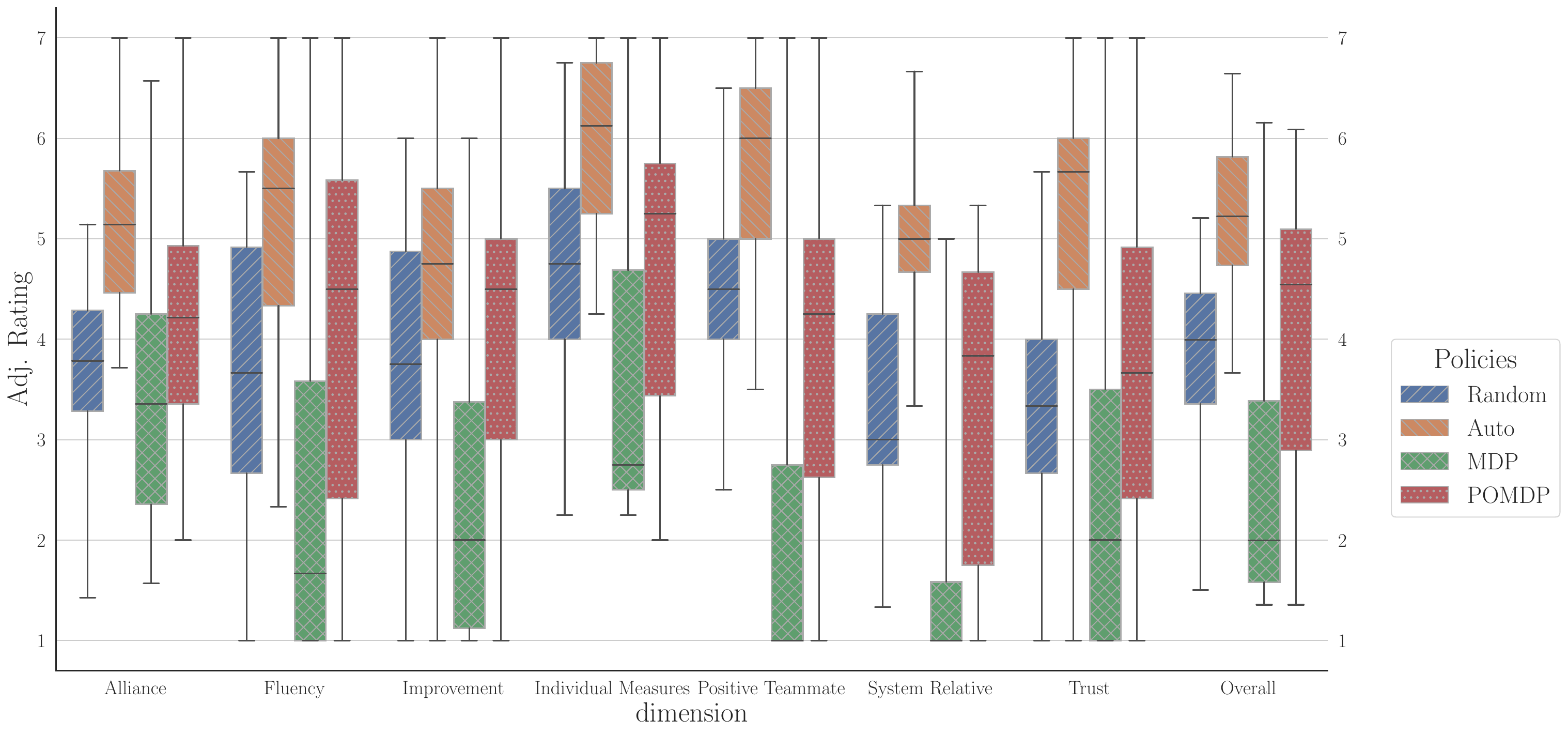}
\caption{Boxplot with the distributions of the Fluency test by interaction control policy and dimension. The Fixed Automatic policy (FA or Auto) is perceived as the one that leads to the most fluent interaction, while the MDP adaptive control policy is perceived as the least fluent.}
\label{fig:fluency}
\end{figure}

The results of the NASA-TLX questionnaire are displayed in Figure \ref{fig:nasatlx}.
\FloatBarrier
 The results of the Fluency questionnaire are displayed in Figure \ref{fig:fluency}.
